\documentclass[aps,prb,10pt,twocolumn,amsfonts,showpacs,longbibliography,citeautoscript,superscriptaddress]{revtex4-1}

\usepackage{color}
\usepackage{bm}
\usepackage{bbold}
\usepackage{amsmath}
\usepackage{amssymb}
\usepackage{epsfig}
\usepackage{verbatim}
\usepackage[T1]{fontenc}
\usepackage{array}
\usepackage{feynmp-auto}
\usepackage{nicefrac}

\usepackage{tikz}
\usetikzlibrary{calc,arrows}
\def\centerarc[#1](#2)(#3:#4:#5){ \draw[#1] ($(#2)+({#5*cos(#3)},{#5*sin(#3)})$) arc (#3:#4:#5); }

\newcommand{\be}{\begin{equation}}
\newcommand{\ee}{\end{equation}}
\newcommand{\bw}{\begin{widetext}}
\newcommand{\ew}{\end{widetext}}
\newcommand{\bea}{\begin{eqnarray}}
\newcommand{\eea}{\end{eqnarray}}
\newcommand{\la}{\langle}
\newcommand{\ra}{\rangle}
\newcommand{\dg}{^\dagger}
\newcommand{\p}{\partial}
\newcommand{\rd}{{\rm d}}
\newcommand{\s}{\sigma}

\newcommand{\re}{\text{Re}}

\def\nn{\nonumber\\}

\def\vec#1{{\mathbf #1}}

\usepackage[colorlinks=true,bookmarks=false,citecolor=blue,linkcolor=blue,hyperfootnotes=true,urlcolor=blue]{hyperref}

\allowdisplaybreaks

\begin{document}

\title{Ladder-like optical conductivity in the spin-fermion model}

\author{Laura Classen}
\affiliation{Condensed Matter Physics \& Materials Science Division, Brookhaven National Laboratory, Upton, NY 11973-5000, USA}

\author{Neil J. Robinson}
\affiliation{Institute for Theoretical Physics, University of Amsterdam, Science Park 904, 1098 XH Amsterdam, The Netherlands}

\author{Alexei M. Tsvelik}
\affiliation{Condensed Matter Physics \& Materials Science Division, Brookhaven National Laboratory, Upton, NY 11973-5000, USA}

\date{\today}

\begin{abstract} 
In the nested limit of the spin-fermion model for the cuprates, one-dimensional physics in the form of half-filled two-leg ladders emerges. We show that the renormalization group flow of the corresponding ladder is towards the d-Mott phase, a gapped spin-liquid with short-ranged d-wave pairing correlations, and reveals an intermediate SO(5)$\times$SO(3) symmetry. We use the results of the renormalization group in combination with a memory-function approach to calculate the optical conductivity of the spin-fermion model in the high-frequency regime, where processes within the hot spot region dominate the transport. We argue that umklapp processes play a major role. For finite temperatures, we determine the resistivity in the zero-frequency (dc) limit. Our results show an approximate linear temperature dependence of the resistivity and a conductivity that follows a non-universal power law. A comparison to experimental data supports our assumption that the conductivity is dominated by the antinodal contribution above the pseudogap.
\end{abstract}

\maketitle

\section{Introduction}

The interplay between magnetism and superconductivity seems to be at the heart of high-temperature superconductivity in various families of materials~\cite{akhavan2006interplay}. It becomes particularly interesting when, in special parameter regimes, the mutual reinforcement of instabilities in magnetic and pairing channels impedes a definite description of the observed phase. A prime example for such a situation is the anomalous behavior of underdoped cuprates, where the mysterious pseudogap and strange-metal phases appear~\cite{timusk1999pseudogap,norman2005pseudogap,lee2006doping,hussey2008phenomenology,hashimoto2014energy,fradkin2015colloquium}.

Among many approaches to the problem of the underdoped regime, let us focus on two phenomenological ones. The first is based on the spin-fermion model~\cite{abanov2003quantum}, while the second suggests an analogy between two-dimensional (2D) doped Mott insulators and one-dimensional (1D) ladders, as put forward by Dagotto and Rice~\cite{dagotto1996suprises}. Both approaches have been quite successful, which gives rise to the question of whether they can be brought, so to speak, to a common denominator?

One of the main motivations for invoking the physics of two-leg ladders in the context of the cuprates is that these systems represent the first step away from 1D towards 2D. Despite moving towards 2D, in ladders one still has well-controlled access to the strong coupling regime via powerful nonperturbative techniques peculiar to 1D~\cite{giamarchi2004quantum,gogolin2004bosonization,james2018nonperturbative}. Of particular relevance to the pseudogap phase, ladder physics provides a simple mechanism for the formation of both a spin gap and superconducting pairing, with the two appearing simulataneously in one of the phases of undoped fermionic two-leg ladders. This so-called d-Mott phase describes a Mott-insulating spin liquid with short-ranged d-wave pairing correlations, which upon doping develop into (quasi-)long-ranged superconductivity~\cite{balents1996weakcoupling,lin1998exact}.

Alternatively, the spin-fermion model is a fully 2D theory~\cite{abanov2003quantum}. It is based on the assumption that the anomalous behavior in the underdoped regime of the cuprates is caused by the vicinity to a magnetic quantum critical point. It approaches the anomalous phase from the high-doping side, where the electronic state with a large Fermi surface becomes unstable due to interactions in the spin channel. Below the critical point, this leads to the antiferromagnetic state. At the same time, the exchange of paramagnetic spin-fluctuations in the nonmagnetic phase provides the pairing glue for d-wave superconductivity. However, above antiferromagnetic and superconducting transitions, the physics is driven by the interaction of electrons with collective spin-excitations (paramagnons) leading to an incoherent quantum-critical regime. An essential ingredient of the spin-fermion model is the existence of so-called hot spots on the Fermi surface, which are connected by singular spin modes. 

These two paradigms of cuprate physics appear, at first glance, to be unrelated. Yet this is not the case. In  previous work by one of the authors~\cite{tsvelik2017ladder} the spin-fermion model was studied in a limit where the Fermi surface becomes nested around the hot spots. It was argued that such a situation emerges self-consistently when the bare interactions are sufficiently strong, since the nesting leads to the gap formation and the system benefits energetically from it. As such, the size of the nested patches is determined, self-consistently, by the competition between the size of the gap and the deviation of the bare Fermi surface from the nesting condition.  The paramagnon exchange interaction inside and between these flat, nested patches is singular in momentum space and effectively decouples them from the rest of the Fermi surface, forming an effective half-filled two-leg ladder in momentum space. Thus the physics of 1D ladders is brought to bear on the 2D spin-fermion model.

For superconductivity to emerge in this formulation of the spin-fermion model, it is essential that the resulting ladder lies in the region of parameter space describing the d-Mott phase. This was explicitly demonstrated in Ref.~\onlinecite{tsvelik2017ladder}, so providing a mapping that unifies these two well-known approaches to the anomalous phases of the cuprates. Within this picture, the excitation gap of the d-Mott phase relates to the pseudogap and the cooperon excitations of the ladder play the role of preformed pairs. These effects appear around the original hot spots, i.e. in the antinodal region when considering the cuprate Fermi surface below optimal doping. On the other hand, electrons away from the hot spots, in the nodal region, are not subject to the mapping to ladders and remain Fermi-liquid like. It was suggested that a coupling between the ladder degrees of freedom and the nodal electrons promotes the pairing correlations of the d-Mott phase to true superconductivity~\cite{tsvelik2017ladder}. 
  
In other recent work~\cite{rice2017umklapp} this ladder paradigm was exploited to develop a qualitative description of the transport in underdoped cuprates. According to this description, the electrons in the underdoped phase of the cuprates can be separated into two weakly coupled liquids. The first is Fermi-liquid like, describing the nodal quasi-particles. The second, however, describes electrons in the vicinity of the hot spots as a quasi-1D strongly correlated liquid.  These two liquids have very different effective dimensionalities, and hence display very different physics. The parameters of the phenomenological model in Ref.~\onlinecite{rice2017umklapp} were chosen such that the high temperature transport is dominated by contributions from the hot spots, which at these temperatures can be described as a Luttinger liquid. An essential characteristic of this quasi-1D liquid is that vertex corrections play an important role in the diagrammatic expansion, leading to a liquid that has excellent transport properties whilst having no coherent quasiparticles. This helps to explain the drastic difference between quasiparticle lifetimes and transport lifetimes observed in experiments~\cite{timusk1999pseudogap,hussey2008phenomenology}. The finite conductivity of electrons in the vicinity of the hotspots arises from umklapp scattering, whose influence grows with decreasing temperature until, eventually, spectral gaps form in the quasi-1D liquid. At temperatures below this gap formation, the transport becomes dominated by the nodal quasiparticles. This scenario explains some of the characteristic features of transport in the underdoped (pseudogap) regime of the cuprates, including the different transport times observed in the longitudinal and Hall conductivities~\cite{rice2017umklapp}. The latter of these arises only from the curved region of the Fermi surface~\cite{ong1991geometric}, and is thus governed by the nodal quasiparticle Fermi-liquid-like transport time.

In this work  we take a more quantitative approach and compute the contribution of the strongly-correlated hot spots to the optical conductivity. This provides a self-consistency test of our theory and, more importantly, a direct link to experiment. The optical properties of a material reveal fundamental information about its excitation spectrum, including the aforementioned pseudogap in the underdoped cuprates, which breaks the Fermi surface into disconnected nodal and antinodal sections~\cite{puchkov1996pseudogap,timusk1999pseudogap,norman2005pseudogap,lee2006doping,hussey2008phenomenology,hashimoto2014energy,fradkin2015colloquium}. Consistent with our approach, experimental results observe two different components to the optical response, one being Fermi-liquid-like and the other being incoherent~\cite{pines1997theory,mirzaei2013spectroscopic}.

Previous calculations of the optical conductivity from the spin-fermion model have been performed in the non-nested situation~\cite{moraghebi2002optical,abanov2003quantum,haule2007optical,metlitski2010quantum,hartnoll2011quantum,chubukov2014optical},
while calculations of the optical conductivity in fermionic two-leg ladder systems have considered either doped situations~\cite{hayward1996optical,essler2007dynamical}, or the integrable, infrared limit of the half-filled case~\cite{konik2001exact}. In the first two cases,  umklapp processes are irrelevant away from half filling or for a non-nested Fermi surface, which generally has consequences for (dynamical) correlation functions. The integrable limit of the half-filled ladder, on the other hand, is due to an emergent higher symmetry in the far infrared~\cite{lin1998exact} and potentially cannot be reached for realistic values of the bare couplings or on experimentally-relevant energy scales. In this sense, we access a non-traditional regime for the optical conductivity with respect to both models. Another view is that the comparison of observables calculated within this ladder approach with the ones obtained from the self-consistent, nested solution of the spin-fermion model will allow one to see if the mapping between both~\cite{tsvelik2017ladder} is really a unification or if it rather reveals additional 1D-type physics that could mask the spin-fermion behavior.

In addition to the motivation above, we would like to emphasize that both the spin-fermion model and ladder models have many applications beyond the high temperature cuprate superconductors. The spin-fermion model represents a general theory for an antiferromagnetic quantum critical point, which, for example, can also occur in heavy-fermion materials~\cite{stewart1984heavyfermion,si2010heavy}. Ladders can also appear as key structural units in other materials (see, e.g., the introduction of Ref.~\onlinecite{carr2013spinful}). Although our starting point is formally a ladder based on a momentum space decomposition, our calculation remains applicable for systems built of real-space ladders. In fact, there are even ladder compounds in the cuprate familiy~\cite{dagotto1999experiments}, which provide another interesting connection between unconventional superconductivity in 1D and 2D.

To calculate the optical conductivity, we use a perturbative memory function formalism~\cite{gotze1972homogeneous,giamarchi1991umklapp} in combination with the one-loop renormalization group (RG). This allows us to accurately access the high-energy regime, at frequencies or temperature significantly above the excitation gap. Such an ``RG-improved'' perturbation theory has been previously used to study the optical conductivity in the sine-Gordon model, where it was shown to be accurate to a comparable level to the exact solution at intermediate-to-high energy
\cite{controzzi2001optical,controzzi2000dynamical}. The combination of the RG with a perturbative determination of the memory function has also been used to explain conductivity measurements in the ladder compound Sr$_{14-x}$Ca$_x$Cu$_{24}$O$_{41}$~\cite{byrne2002role}. Our calculation appears similar, with the important difference that umklapp processes are relevant in our case and we account for the scale-dependence of the Luttinger parameters characterizing the bosonized version of the ladder. As in the previous work~\cite{rice2017umklapp}, we assume that at high energies/temperatures the processes within the antinodal regions dominate the transport. This assumption has allowed Ref.~\onlinecite{rice2017umklapp} to get a good qualitative fit to the experimentally-measured temperature dependence of the dc resistivity in underdoped cuprates, and helps to justify our focus on this region in the mapping from the spin-fermion model to the ladder. 

We find that umklapp processes play a major role in the high-frequency behavior of the optical conductivity, leading to a high-frequency tail that falls off like a power law $\omega^{-\alpha}$ with a non-universal exponent $\alpha$. If we consider finite temperatures and the zero-frequency limit in turn, we find a resistivity that appears to decrease linearly as the temperature is lowered, before diverging at low temperatures. This confirms the assumption in Ref.~\onlinecite{rice2017umklapp} and, as has been argued there, when this is combined with the contribution from nodal quasiparticles, the divergence is regularized and substituted by Fermi-liquid scaling of the resistivity at low temperatures. Such behavior is qualitatively consistent with that observed in experiments, see Ref.~\onlinecite{mirzaei2013spectroscopic}.
 
In the following section, we explain the correspondence between the hot spot regions of the spin-fermion model and the half-filled, two-leg ladder. In Sec.~\ref{sec:RG}, we argue how the RG flow reveals the d-Mott phase, even before the integrable limit is reached in the far infrared. We present our results for the optical conductivity in Sec.~\ref{sec:oc} and discuss our conclusions in Sec.~\ref{sec:conc}.

\section{Ladder physics in the spin-fermion model}

\begin{figure}
  \includegraphics[width=0.4\textwidth]{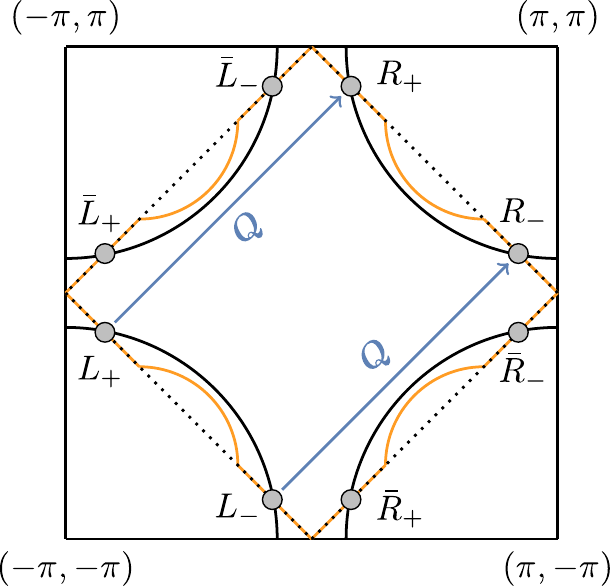}
  \caption{Solid black lines show the non-interacting Fermi surface, which intersects the magnetic Brillouin zone (dotted lines) at eight isolated points, the hot spots (grey spots). The hot spots are nested via the antiferromagnetic wave vector $\bm{Q} = (\pm\pi,\pm\pi)$. Generically the velocities at opposing hotspots are not equal and opposite. In the presence of spin-fermion interaction, we consider a Fermi surface that has deformed in the vicinity of the hot spots to increased nesting (shown exageratedly in orange) and thus lower the overall energy through the opening of spectral gaps.}
\label{Fig:SpinFermionBZ}
\end{figure}

The formal link between the physics of the hot spots in the spin-fermion model and half-filled, two leg ladders is provided by recent work of one of the authors~\cite{tsvelik2017ladder}. We remind the reader of the basic ideas of this approach here. The spin-fermion model~\cite{abanov2003quantum} describes electrons interacting with soft spin excitations (paramagnons) that emerge in the vicinity of an antiferromagnetic quantum critical point. The low-energy Lagrangian features electrons $\psi_\s(\bm{k})$ with a large Fermi surface and collective spin excitations $\bm{S}_{\bm{q}}$: 
\begin{align}
{\cal L} &= \sum_{\bm{k}}  \psi\dg_{\s}(\bm{k})(i\omega-\epsilon_{\bm{k}}) \psi_{\s}(\bm{k}) + \frac12 \sum_{\bm{q}} \bm{S}_{\bm{-q}} \chi^{-1}(\bm{q}) \bm{S}_{\bm{q}} \nn
& \quad + g\sum_{\bm{k},\bm{q}} \psi\dg_{\alpha}(\bm{k}+\bm{q}) \bm{\s}_{\alpha\beta} \psi_\beta(\bm{k}) \cdot \bm{S}_{\bm{q}}\,.
\end{align}
Here $\bm{\s}$ is the vector of Pauli matrices. The spin susceptibility $\chi(\bm{q})$ is
\be
\chi(\bm{q}) = \frac{\chi_0}{ 1 + \xi^2\big( \bm{Q} - \bm{q}\big)^2},
\ee
where $\bm{Q} = (\pm\pi,\pm\pi)$ are the antiferromagnetic wave vectors that connect hotspots in the antinodal regions of the Fermi surface (see Fig.~\ref{Fig:SpinFermionBZ}), and $\xi$ is the magnetic correlation length. The Fermi surface associated with the dispersion relation $\epsilon_{\bm{k}}$ is shown (with exaggerated nesting deformation) in Fig.~\ref{Fig:SpinFermionBZ} in orange.

The spin-fermion model has been intensively studied~\cite{abanov2000spinfermion,abanov2003quantum,abanov2004,metlitski2010quantum,berg2012,shouvik2015,schattner2016,maierstrack2016,schlief2017exact,lunts2017} but, despite remarkable progress, the full RG equations for a finite number of hotspots in 2D have not been solved, even at one loop. RG calculations show a logarithmic increase of the coupling constant and an increased tendency to nesting at the hot spots~\cite{abanov2000spinfermion}. These calculations also show a decrease in the dynamical exponent $z$~\cite{metlitski2010quantum}.  Recently, Schlief, Lunts and Lee argued that the theory with $z=1$ is self-consistent with the result that the coupling flows to zero. At the same time, they claim that the theory remains strongly coupled because the dimensionless coupling is of order one~\cite{schlief2017exact}. In our calculations we assume that there is some critical value of the bare coupling constant above which the system scales to perfect nesting and strong coupling (meaning a non-zero, dimensionful coupling constant). We justify this by the fact that gap creation is energetically advantageous and, if the gap is sufficiently large, it can win over the losses in the kinetic energy caused by the imperfect nesting. This is similar to the scenario envisaged by Rice to explain antiferromagnetism at incommensurate fillings in Co alloys~\cite{rice1970bandstructure}.

In this limit, we focus on the physics in the vicinity of the hot spots. We neglect any contribution from the nodal regions (i.e., electrons away from the hotspots) but comment on potential modifications of our results below. Doing so allows us to project onto the hot spots via\cite{tsvelik2017ladder}
\begin{align}
&R_{\s a}(x)=\frac{1}{\sqrt{2\pi}}\int \rd k_\|\, \psi(\vec k_R^a+k_\| \vec e) e^{ik_\|x}, \nn
&L_{\s a}(x)=\frac{1}{\sqrt{2\pi}}\int \rd k_\|\, \psi(\vec k_L^a+k_\| \vec e) e^{ik_\|x},
\end{align}
where $\vec k_{R/L}^a$  denotes the  coordinates of the hot spots, $a=+,-$, and $\vec e=(1,1)/\sqrt{2}$ (see Fig.~\ref{Fig:SpinFermionBZ}). Integrating out the paramagnons under the assumption that the correlation length remains finite, we arrive at the low-energy effective Lagrangian density\cite{tsvelik2017ladder} of a 1D fermionic two-leg ladder without interchain hopping, but with interchain interactions determined by the spin-fermion coupling
\begin{align}
  {\cal L}=& R\dg_{\s a}\big(\p_\tau-iv\p_x\big)R_{\s a} + L\dg_{\s a}\big(\p_\tau+iv\p_x\big)L_{\s a} \nonumber \\
  &- \frac{\gamma}{2} \Big(R\dg_{\alpha a} \vec{\s}_{\alpha\beta}L_{\beta a} + L\dg_{\alpha a} \vec{\s}_{\alpha\beta}R_{\beta a}\Big)\nonumber\\
  & \qquad \times \Big(R\dg_{\gamma b} \vec{\s}_{\gamma\delta}L_{\delta b} + L\dg_{\gamma b} \vec{\s}_{\gamma\delta}R_{\delta b}\Big),
\end{align}
where $\gamma\sim g^2\chi_0/\xi$ and $v$ is the Fermi velocity, which only has a component perpendicular to the Fermi surface in the flat, nested limit that we consider. The model possesses  U(1)$\times$U(1)$\times$SU(2)$\times\mathbb{Z}_2$ symmetry as a result of charge conservation within each pair of patches, spin conservation, and the symmetry under exchange of the two pairs of patches.

\begin{figure}[t]
\includegraphics[width=0.48\textwidth]{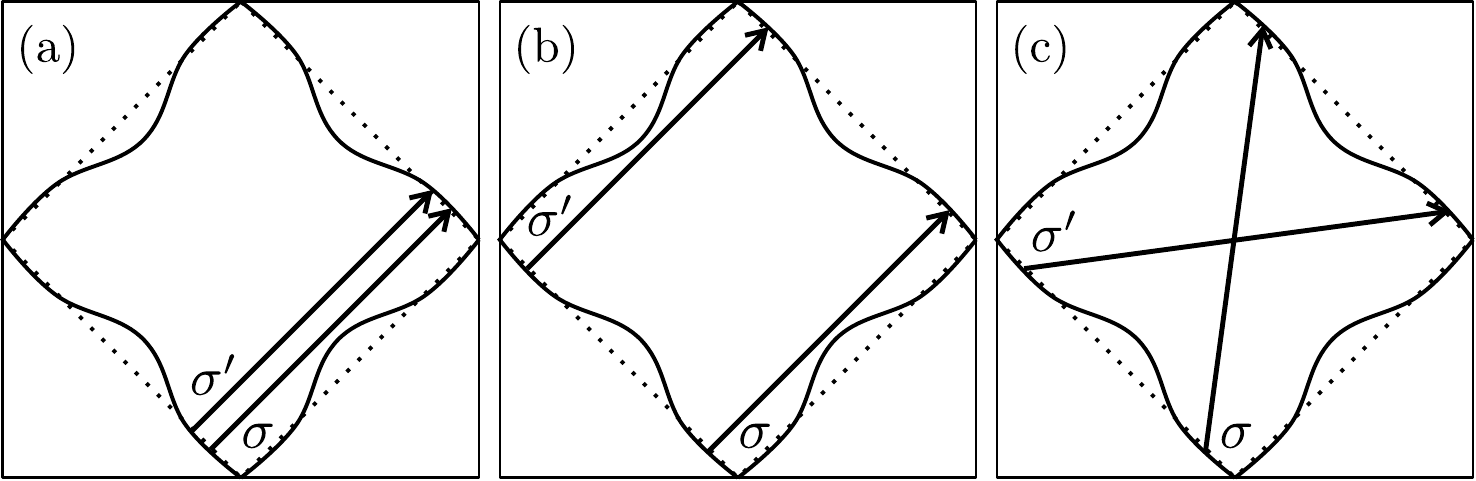}
\caption{Umklapp processes of the deformed 2D Fermi surface that correspond to the relevant umklapp terms of the half-filled ladder. They are proportional to the $\cos(\sqrt{4\pi}\Phi_c)$-terms in Eq.~\eqref{eq:bosonized} with coupling (a) $g_{cf}$ and spin $\sigma=\sigma'$, (b) $g_{cs}$ for $\sigma=\sigma'$ and $g_{csf}$ for $\sigma\neq\sigma'$, and (c) $g_{\overline{csf}}$ with $\sigma\neq\sigma'$.}
\label{fig:Umklapps}
\end{figure}

With the help of bosonization~\cite{gogolin2004bosonization,giamarchi2004quantum}, we can express the Lagrangian in terms of four scalar fields via
\begin{align}
& R_{\s p}=\frac{\kappa_{\s p}}{\sqrt{2\pi a_0}}e^{i\sqrt{\pi}(\varphi_c + \s\varphi_s + p\varphi_f + \s p\varphi_{sf})}, ~~ \s, p = \pm 1, \nonumber\\
&  L_{\s p}=\frac{\kappa_{\s p}}{\sqrt{2\pi a_0}}e^{-i\sqrt{\pi}(\bar\varphi_c + \s\bar\varphi_s + p\bar\varphi_f + \s p\bar\varphi_{sf})}, \label{F}
\end{align}
with small-distance regularization $a_0$ and Klein factors $\{\kappa_{\s a},\kappa_{\s' b}\}=2\delta_{\s\s'}\delta_{ab}$. In our convention, $\kappa_{\s a}\kappa_{\s-a}\kappa_{-\s-a}\kappa_{-\s a}=1$ and the correlators of the bosonic fields satisfy
\begin{align}
  & \la \varphi_a(x,\tau) \varphi_b(0,0) \ra = \frac{\delta_{a,b}}{4\pi} \log\left( \frac{a_0}{\tau + ix/v}\right) \\
  & \la \bar\varphi_a(x,\tau) \bar\varphi_b(0,0) \ra = \frac{\delta_{a,b}}{4\pi}\log\left(\frac{a_0}{\tau - ix/v} \right).
\end{align}
Here $\tau$ is imaginary time. For convenience we are working with bosonic fields describing charge ($c$), spin ($s$), flavor ($f$) and spin-flavor ($sf$) degrees of freedom, and it will also be useful to introduce non-chiral fields $\Phi_{a}=\varphi_a + \bar\varphi_{a}$ and their duals $\Theta_{a}=\varphi_{a} - \bar\varphi_{a}$.

With these definitions at hand, the bosonized version of the Lagrangian density becomes
\begin{widetext}
\begin{align}
  {\cal L}  =& \sum_{\mu=c,f,s,sf} \frac{1}{2K_\mu}\left[ \frac{1}{v}(\partial_\tau\Phi_\mu)^2 + v(\partial_x\Phi_\mu)^2\right] + \frac{2g_{ssf}}{(\pi a_0)^2} \cos( \sqrt{4\pi} \Phi_s)\cos(\sqrt{4\pi} \Phi_{sf}) - \frac{g_{cf}}{(\pi a_0)^2} \cos( \sqrt{4\pi} \Phi_c)\cos(\sqrt{4\pi} \Phi_f )\nonumber\\
&  + \frac{1}{(\pi a_0)^2} \Big[ \cos(\sqrt{4\pi} \Phi_c) + \cos(\sqrt{4\pi}\Phi_f) \Big] \Big[ g_{cs} \cos(\sqrt{4\pi}\Phi_s) -g_{csf} \cos(\sqrt{4\pi}\Phi_{sf}) + 2 g_{\overline{csf}} \cos(\sqrt{4\pi}\Theta_{sf}) \Big] \label{eq:bosonized}
\end{align}
with $1/K_{c(f)}= 1+g_c/(2\pi v)$ and $1/K_{s(sf)}=1-g_{s(sf)}/(2\pi v)$. We list the bare values of the couplings below, see Eqs.~\eqref{eq:bare}. As we consider a deformed Fermi surface with increased nesting about the hot spots, terms proportional to $\cos(\sqrt{4\pi}\Phi_c)$ appear in the Lagrangrian~\eqref{eq:bosonized}, which derive from the umklapp processes shown in Fig.~\ref{fig:Umklapps} and are marginally relevant at half-filling. They appear on the same footing as $\cos(\sqrt{4\pi}\Phi_f)$ terms, because there is a symmetry with respect to $\Phi_c \leftrightarrow\Phi_f$ at half-filling. In contrast to the $\cos(\sqrt{4\pi}\Phi_c)$ terms, however, the $\cos(\sqrt{4\pi}\Phi_f)$ terms survive finite doping away from half filling.

To proceed further, we need to derive how the couplings flow under the RG. To do so, it is convenient to refermionize the Lagrangian~\eqref{eq:bosonized} and to this end we define four Majorana fermions (organized into a singlet and a triplet) for the spin degrees of freedom $\xi_i$ ($i=0,\ldots,3$) and four Majorana fermions that characterize the charge sector $\eta_a$ and $\lambda_a$ ($a=c,f$). We refermionize according to the identities: 
\begin{align}
  R_s&=\frac{1}{\sqrt 2}(\bar\xi_1 + i\bar\xi_2) = \frac{\kappa_{s}}{\sqrt{2\pi a_0}}e^{i\sqrt{4\pi}\varphi_s} , \quad
  &L_s = \frac{1}{\sqrt 2}( \xi_1 + i\xi_2) = \frac{\kappa_{s}}{\sqrt{2\pi a_0}}e^{-i\sqrt{4\pi}\bar\varphi_s}, \nn
R_{sf}&=\frac{1}{\sqrt 2}(\bar\xi_0 + i\bar\xi_3) = \frac{\kappa_{sf}}{\sqrt{2\pi a_0}}e^{i\sqrt{4\pi}\varphi_{sf}}, \quad
 &L_{sf} = \frac{1}{\sqrt 2}( \xi_0 + i\xi_3) = \frac{\kappa_{sf}}{\sqrt{2\pi a_0}}e^{-i\sqrt{4\pi}\varphi_{sf}},\nn
R_a &= \frac{1}{\sqrt 2}(\bar\eta_{a} + i\bar\lambda_{a}) = \frac{\kappa_{a}}{\sqrt{2\pi a_0}}e^{i\sqrt{4\pi}\varphi_{a}},  \quad
&L_a = \frac{1}{\sqrt 2}( \eta_{a} + i\lambda_{a}) = \frac{\kappa_{a}}{\sqrt{2\pi a_0}}e^{-i\sqrt{4\pi}\varphi_{a}}. 
\end{align}
Here $\{\kappa_a,\kappa_b\} = 2\delta_{a,b}$ are new Klein factors. As one can see, these new fermions are nonlocal with respect to the original ones~\eqref{F}. In terms of these new Majorana fermions, the Lagrangian density reads:  
\begin{align}
 {\cal L} &= \sum_{i=0}^3\left[\bar\xi_i(\p_\tau-iv\p_x)\bar\xi_i + \xi_i(\p_\tau+iv\p_x)\xi_i \right]  + \sum_{a=c,f}\left[ \bar\eta_a(\p_\tau-iv\p_x)\bar\eta_a + \eta_a(\p_\tau+iv\p_x)\eta_a  \right]  \nonumber\\
& + \sum_{a=c,f}\left[ \bar\lambda_a(\p_\tau-iv\p_x)\bar\lambda_a + \lambda_a(\p_\tau+iv\p_x)\lambda_a  \right] + g_c\sum_{a=c,f}\bar\eta_{a}\eta_{a}\bar\lambda_{a}\lambda_{a} +g_{cf}(\bar\eta_{c}\eta_{c} +\bar\lambda_{c}\lambda_{c})(\bar\eta_{f}\eta_{f} + \bar\lambda_{f}\lambda_{f}) \nonumber\\
& - \sum_{a=c,f}(\bar\eta_{a}\eta_{a} +\bar\lambda_{a}\lambda_{a})(g_{cs,+}\bar\xi_b\xi_b+ g_{cs,-}\bar\xi_0\xi_0) -g_{s,+}\sum_{a>b}(\bar\xi_a\xi_a) (\bar\xi_b\xi_b) - g_{s,-}(\bar\xi_a\xi_a) (\bar\xi_0\xi_0).
\label{eq:majoranaexp}
\end{align}
\end{widetext}
There are six couplings in total, with the bare values
\be\label{eq:bare}
g_{c}^0 =g_{cf}^0=-g_{cs,-}^0= 3\gamma, \quad  g_{cs,+}^0 =g_{s,+}^0=g_{s,-}^0= \gamma.
\ee
These couplings are related to those of the bosonized Lagrangian density, Eq.~\eqref{eq:bosonized}, through
\begin{align}
g_s&=g_{s+}, ~~ g_{sf}=g_{s-}, ~~ g_{ssf}=\frac{g_{s+}+g_{s-}}{4},\nonumber\\
g_{cs}&=g_{cs+}, ~~ g_{csf}=-\frac{g_{cs+}+g_{cs-}}{2}, ~~ g_{\overline{csf}}=\frac{g_{cs+}-g_{cs-}}{4}. \nonumber
\end{align}
Although the couplings $g_{s+}$ and $g_{s-}$ satisfy $g_{s+}= g_{s-}$ at the bare level, this can be broken under the RG flow. As a result a new interaction term, which is proportional to the coupling $g_{\overline{ssf}}=(g_{s+}-g_{s-})/2$ will be generated, $g_{\overline{ssf}}\cos(\sqrt{4\pi}\Phi_s) \cos(\sqrt{4\pi}\Theta_{sf})$.

In Eq.~\eqref{eq:majoranaexp}, an SO(5)$\times $SO(3) symmetry becomes apparent.\footnote{The restriction to the special orthogonal group in O(N)$\cong$SO(N)$\times \cal{Z}_2$ is not directly obvious, but $\cal{Z}_2$ transformations of the Majorana fermions are unphysical in the original fermions~\cite{lin1998exact}.$~$} This can be made very explicit by collecting $\eta_{c,f},\lambda_{c,f}, \xi_0$ fermions into a quintet denoted by $\chi_a$ ($a= 1,2,\ldots,5$) and $\xi_{1,2,3}$ fermions into a triplet. The interaction term then reduces to the (obviously) symmetric form  
\begin{align}
 V =& - g_c\sum_{a>b}(i\bar\chi_a\chi_a)(i\bar\chi_b\chi_b) + g_{s,+}\sum_{a>b}(i\bar\xi_a\xi_a)(i\bar\xi_b\xi_b)  \nonumber\\
 & + g_{cs,+}(i\bar\chi_a\chi_a)(i\bar\xi_b\xi_b). \label{eq:symmetry}
 \end{align}

As described in the next section, all of these excitations develop a gap. It can be reasonably assumed (see Ref.~\onlinecite{tsvelik2017ladder} and below) that the lowest excitations of this theory are the same as those of the SO(8) Gross-Neveu model; in particular, there are eight excitations having nonzero overlap with the Majorana fermions~\cite{lin1998exact,konik2001exact}. They are approximately split into a triplet and a quintet, with the triplet ones being related to $S=1$ magnetic excitations, and the quintet containing, among other excitations, a gapped $2e$-charged magnetic singlet -- the so-called cooperon. This is different from previously studied SO(5) symmetry that combines spin and pairing excitations into a quintet\cite{lin1998exact,scalapino1998,shelton1998,arrigoni99,demler2004}.

\section{RG flow and $\textrm{d}$-Mott phase}
\label{sec:RG}

\begin{figure}[t]
\includegraphics[width=0.48\textwidth]{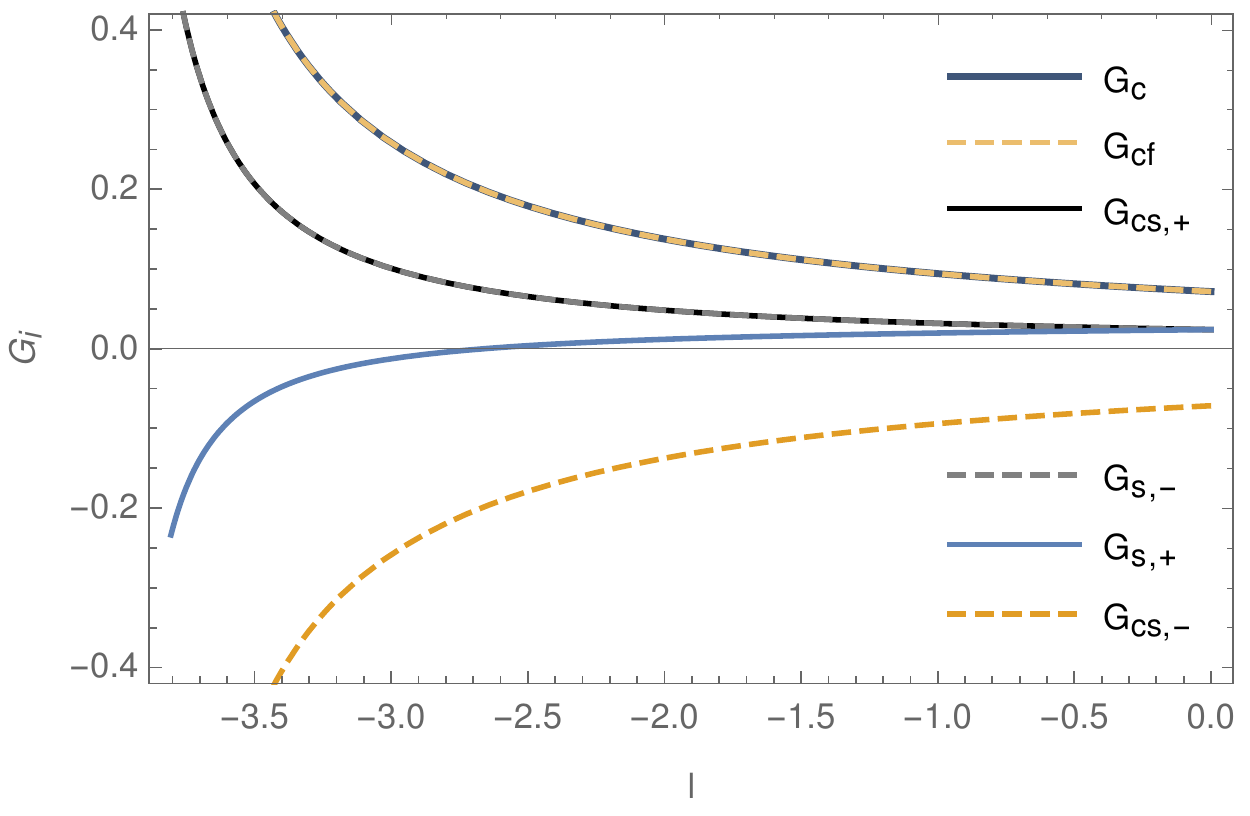}
\includegraphics[width=0.48\textwidth]{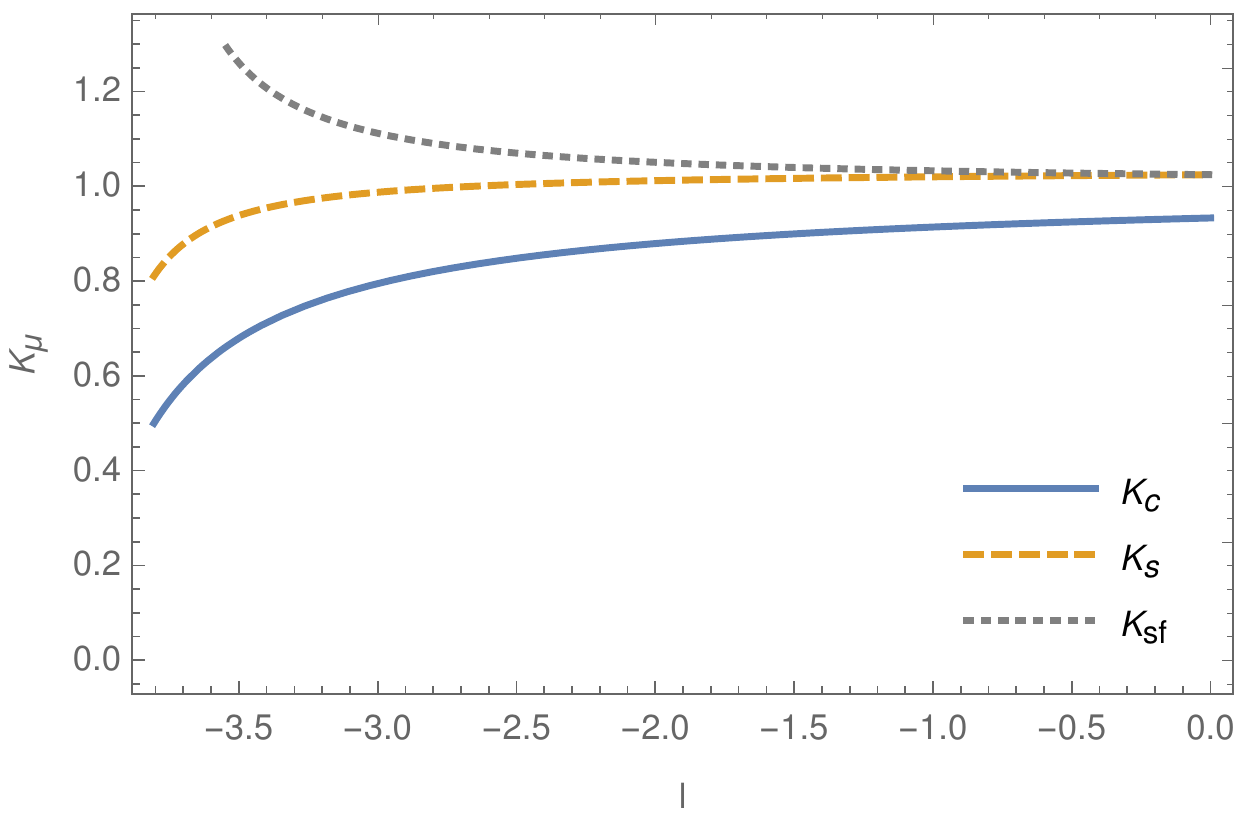}
\caption{Flow of the couplings and Luttinger parameters for $\gamma=0.15v$. All couplings flow to strong coupling, and we stop the RG flow when one $G_\mu$ becomes of order one. The initial SO(5)$\times$ SO(3) symmetry is maintained or emerges at intermediate scales when we perturb the initial conditions.}
\label{fig:flow}
\end{figure}

The RG equations for different two-leg ladder models have been studied in many previous works\cite{khveshchenko1994spingap,balents1996weakcoupling,lin1998exact,byrne2002role,controzzi2005excitation,jaefari2012pairdensitywave,robinson2012finite}. Although ladder models are frequently considered in the limit of strong interchain tunneling, a strong-weak tunneling duality allows one, in principle, to relate the low-energy effective action of the two limits with each other. With this in mind, we present the RG equations here again for the model of Eq.~\eqref{eq:majoranaexp} to explicitly show the relation between the ladder physics in the spin-fermion model and previous results for two-leg ladders. We do not assume the enlarged SO(5)$\times$ SO(3) symmetry holds from the beginning by allowing all six couplings to be different. The RG equations have the following form
\begin{align}
\frac{\rd}{\rd l} G_{c} &= - 2G_{cf}^2- 3G^2_{cs,+} - G_{cs,-}^2, \nonumber\\
\frac{\rd}{\rd l} G_{cf} &= - 2G_{cf}G_c - 3G^2_{cs,+} - G_{cs,-}^2\nonumber\\
\frac{\rd}{\rd l}{G}_{cs,-} &= -(G_{c} + 2G_{cf})G_{cs,-} + 3 G_{s,-}G_{cs,+} \nonumber\\
\frac{\rd}{\rd l}{G}_{cs,+} &= -\Big( G_{c}+2G_{cf} -2G_{s+}\Big)G_{cs,+} + G_{s-}G_{cs,-} \nonumber\\
\frac{\rd}{\rd l}{G}_{s-} &=  2G_{s+}G_{s-} +4G_{cs,+}G_{cs,-} \nonumber \\
\frac{\rd}{\rd l}{G}_{s+} &= G^2_{s+} + G^2_{s-} +4G^2_{cs,+}, \nonumber
\end{align}
where $G(l) = g(l)/2\pi v$ and $l = \ln(k/\Lambda)$ characterizes the RG trajectory ($\Lambda$ is the UV cutoff). As is standard, we neglect irrelevant chiral terms that are generated by the RG. This also means that we neglect the renormalization of the velocity, and in particular, the velocities in the different sectors (charge, spin, flavor, spin-flavor) remain the same $v_c=v_f=v_s=v_{sf}=v$.

The flow of the six couplings is shown in Fig.~\ref{fig:flow}. In agreement with previous RG studies of two-leg ladders, the absolute values of all couplings grow. We stop the flow when one of the couplings $G_\mu$ becomes of order one. The system preserves the SO(5)$\times$SO(3) symmetry of the initial conditions, i.e. 
\be
g_c = g_{cf} = - g_{cs,-}, ~~g_{cs,+} = g_{s,-}, \nonumber
\ee
while $g_{s+}$ deviates from the other couplings and changes sign during the flow. We also checked that the system flows to the SO(5)$\times$SO(3) symmetry when we perturb the initial couplings away from it.  In the strong coupling limit, the system scales to the SO(8) Gross-Neveu model, where the absolute values of all couplings are equal~\cite{lin1998exact}. But this happens far beyond the perturbative regime; on all relevant intermediate scales our system can be considered to be SO(5)$\times$SO(3) symmetric. As a result, we cannot take advantage of the integrable point which would be reached with the SO(8) Gross-Neveu model~\cite{lin1998exact}. However, as explained below, the excitations can still be classified with the same quantum numbers as in the SO(8) symmetric case, but with different energies of the triplet and quintet.

The development of a spectral gap is signaled by the flow to strong coupling, where the SO(5)$\times$SO(3)-symmetric interaction takes the form
\bw
\begin{align}
(\pi a_0)^2V\rightarrow &\bigg[\cos(\sqrt{4\pi}\Phi_c) + \cos(\sqrt{4\pi}\Phi_f)\bigg]\bigg[ {g_{cs,+}}\cos(\sqrt{4\pi}\Phi_s) + \frac{g_{cf}+g_{cs,+}}{2}\cos(\sqrt{4\pi}\Theta_{sf})\bigg] \nonumber \\
& -{g_{cf}} \cos(\sqrt{4\pi}\Phi_c)\cos(\sqrt{4\pi}\Phi_f) - \frac{|g_{s,+}|+g_{cs,+}}{2}\cos(\sqrt{4\pi}\Phi_s)\cos(\sqrt{4\pi}\Theta_{sf}) \nonumber \\
                        &\text{+ \bigg( incoherent terms } \propto \cos(\sqrt{4\pi}\Phi_{sf})
                         \bigg). \nonumber
\end{align}
\ew
When the interaction parameters $g_\mu$ growing large, the fields are pinned to the minima of the potential leading to finite masses for charge and spin excitations. Note, that $g_{s,+}$ changed its sign compared to the bare interaction. In the case of strong coupling, there are two vacua with $\Phi_c=\Phi_f=0$, $\Phi_s=\Theta_{sf}=\sqrt{\pi}/2$ or  $\Phi_c=\Phi_f=\sqrt{\pi}/2$, $\Phi_s=\Theta_{sf}=0$.  Quantum numbers of the spectrum are determined by the distance between the minima of the potential, e.g., the different ``topological charges''
\be
Q_\mu\propto \frac{1}{\sqrt{\pi}}\int \rd x\, \partial_x \Phi_\mu, \quad Q_{sf}\propto \frac{1}{\sqrt{\pi}}\int \rd x\, \partial_x \Theta_{sf} \nonumber
\ee
for $\mu\in\{c,s,f\}$ are non-zero for field configurations that approach the different minima at $x=\pm\infty$. Hence, they do not change between the SO(8) and the SO(5)$\times$SO(3) theory, because the position of the minima remains the same also in the SO(8) symmetric case when $g_{cf},g_{cs,+},-g_{s,+}\rightarrow g$~\cite{lin1998exact}. Consequently, we can transfer qualitative conclusions from there.

This strong-coupling fixed point is not the only basin of attraction of the RG equations. Depending on the initial conditions, model~\eqref{eq:majoranaexp} (and likewise the emergent SO(8) theory) can belong to five different phases: a gapless Luttinger liquid (realized when $g_c^{0} <0$ and, as consequence, all couplings scale to zero), a charge-density-wave, a spin-Peierls, an s-Mott or a d-Mott phase. The latter two phases denote Mott-insulating spin-liquid states which have short-ranged pairing correlations with s- or d-wave symmetry. Consistent with expectations from the spin-fermion model, our initial conditions lie in the basin of attraction of the d-Mott phase. That is the RG evolves our system to a phase with spin and charge gaps, and short-ranged d-wave pairing correlations~\cite{tsvelik2017ladder}. This becomes clear when we consider the amplitude of the order parameter that corresponds to d-wave pairing in the spin-fermion model
\begin{align}
\Delta_d&=R_{+\uparrow}L_{-\downarrow}-R_{+\downarrow}L_{-\uparrow} - R_{-\uparrow}L_{+\downarrow}+R_{-\downarrow}L_{+\uparrow} \nonumber\\
        &\propto
          e^{i\sqrt{\pi}\Theta_c}\Big[i\cos(\sqrt{\pi}\Phi_f)\sin(\sqrt{\pi}\Phi_s)\sin(\sqrt{\pi}\Theta_{sf}) \nonumber \\
&\hspace{2cm}+ \sin(\sqrt{\pi}\Phi_f)\cos(\sqrt{\pi}\Phi_s)\cos(\sqrt{\pi}\Theta_{sf}) \Big], \nonumber
\end{align}
which is finite for both the aforementioned vacua. Quasi-long-ranged pairing correlations are expected when the field $\Theta_c$ becomes gapless, e.g., upon doping the ladder away from half-filling. By analogy, we relate the gap and pairing correlations of the d-Mott phase to the pseudogap in the antinodal regions and a tendency towards d-wave pairing, which cannot fully develop because the coupling to the nodal quasiparticles is neglected in our model.

\section{Optical conductivity from RG-improved memory function}
\label{sec:oc}

\subsection{Formalism}

We expect our description to be valid in the perturbative, high-frequency or high-temperature regime before a gap opens in the charge and spin excitation spectrum. Thus, we focus here on two cases: (i) the high-frequency tail of the real part of the optical conductivity; (ii) the high-temperature behavior in the zero-frequency limit. To improve the regime of validity of our approach, we combine a purely perturbative calculation of the optical conductivity with the RG results of the previous section, thereby taking higher orders into account (see, e.g., Ref.~\onlinecite{controzzi2001optical}). This also allows us to detect signatures of the pseudogap in the intermediate frequency regime above the gap opening.

The optical conductivity is determined by the response of the charge sector to an external electric field along the ladder and can be related to the current-current correlation function $\chi(\omega)=\langle\!\langle j;j \rangle\!\rangle$ via $\sigma(\omega)=i[\chi_0+\chi(\omega)]/\omega$, where $\langle\!\langle A;B \rangle\!\rangle=\int\rd t\, \exp(i(\omega+i0^+)t)\la [A(t),B(0)] \ra$, with $\chi_0=2vK_c/\pi$ and we set the electric charge $e=1$. To compute the optical conductivity, we use a memory function approach~\cite{gotze1972homogeneous,giamarchi1991umklapp}
\begin{align}
\sigma(\omega)&=i\chi_0\frac{1}{\omega+M(\omega)}\nn
M(\omega)&=\frac{1}{\chi_0\omega}\Big[ \langle\!\langle F;F \rangle\!\rangle_{\omega} - \langle\!\langle F;F \rangle\!\rangle_{\omega=0}\Big]. \nonumber
\end{align}
$F(t) = [H,j(t)]$ is the commutator of the Hamiltonian and the current operator $j(t)$, which in our theory is given by $j=\sqrt{2/\pi}\p_t \Phi_c$. $F(t)$ is proportional to the scattering potential and to lowest order we can approximate $\langle\!\langle F;F \rangle\!\rangle \approx \langle\!\langle F;F \rangle\!\rangle^0$, i.e. the correlation function is evaluated in the absence of this scattering~\cite{giamarchi1991umklapp}. Evaluating the optical conductivity in terms of an approximated memory function is analogous to calculating particle propagators via approximated self-energies. This approach is valid as long as the couplings of the cosines in Eq.~\eqref{eq:bosonized} remain sufficiently small. 

We see that a finite optical conductivity is due to interaction terms that do not commute with the current operator. In our case, these are exactly the umklapp processes proportional to $\cos(\sqrt{4\pi}\Phi_c)$ in Eq.~\eqref{eq:bosonized} (see also Fig.~\ref{fig:Umklapps}). Let us note that, although we started with a 2D Fermi surface at incommensurate filling, the mapping of the nested Fermi surface in the antinodal regions to the 1D ladder system leads to \textit{commensurate filling} in the effective 1D theory. As a result, arguments~\cite{rosch2000conductivity} that prevent a finite resistivity due to a single, dangerously irrelevant umklapp term do not apply here.

\begin{widetext}
With these preliminaries, we obtain the memory function for our system at finite frequency and temperature
\begin{align}\label{eq:memory_Tw}
 M(\omega,T)\approx\frac{1}{\omega}&
     \left[\sum_{\mu\in\{f,s,sf\}}  c_\mu\,g_{c\mu}^2 \left(\frac{2\pi a_0 T}{v}\right)^{2(K_c+K_\mu)-2} B^2\Big(- \frac{i\omega}{4\pi T}+\frac{K_c+K_\mu}{2},1-K_c-K_\mu \Big) \right. \notag \\
 &\left.+ 4\,{\overline{c}}_{sf}\,\left(\frac{2\pi a_0 T}{v}\right)^{2(Kc+K^{-1}_{sf})-2} B^2\Big(- \frac{i\omega}{4\pi T}+\frac{K_c+K_{sf}^{-1}}{2},1-K_c-K_{sf}^{-2} \Big) - (\omega\rightarrow 0)\right],
\end{align}
where $B(x,y)$ is the Beta function and we define
\begin{align}
  c_\mu=\frac{2K_c}{\chi_0 \pi^4 a_0^2}  \sin\Big(\pi(K_c+K_\mu)\Big), \qquad
  {\overline{c}_{sf}}= \frac{2K_c}{\chi_0 \pi^4 a_0^2} \sin\Big(\pi\big(K_c+K_{sf}^{-1}\big)\Big). \nonumber
\end{align}
We can obtain simplified analytical results in two limits, $T \ll \omega$ and $T\gg \omega$, as discussed below.

\subsubsection{The low temperature limit $T\ll\omega$}
In the low-temperature limit, $T\ll\omega$, the memory function becomes
\begin{align}\label{eq:memory}
M(\omega)\approx\frac{1}{\omega}&\left[\sum_{\mu\in\{f,s,sf\}} c_\mu^0 g_{c\mu}^2 \omega^{2(K_c+K_\mu)-2} + 4\, \overline{c}_{sf}^0\, (g_{\overline{csf}})^2 \omega^{2(Kc+K_{sf}^{-1})-2}\right] \nonumber
\end{align}
with
\begin{align}
  c_\mu^0 & =c_\mu \left(\frac{a_0}{2v}\right)^{2(K_c+K_\mu)-2}
            \exp\Big[i\pi\left(1-K_c-K_\mu\right)\Big] \Gamma^2\Big(1-K_c-K_\mu\Big),\nn 
 {\overline{c}_{sf}^0}&= \overline c_{sf} \left(\frac{a_0}{2v}\right)^{2(K_c+K_{sf}^{-1})-2}
                        \exp\left[i\pi\left(1-K_c-K_{sf}^{-1}\right)\right]
                        \Gamma^2\Big(1-K_c-K_{sf}^{-1}\Big), \nonumber
\end{align}
with $\Gamma(x)$ the Gamma function.

\subsubsection{The high temperature  $T\gg\omega$ limit}
 In the opposite limit, $T\gg\omega$, we find
 \begin{align}
   M(T)=\frac{i}{T}&\left[ \sum_{\mu\in\{f,s,sf\}} c^T_\mu g_{c\mu}^2 \left(\frac{2\pi a_0 T}{v}\right)^{2(K_c+K_\mu)-2} +4\,\overline{c}^T_{sf}\,g_{\overline{csf}}^2 \left(\frac{2\pi a_0 T}{v}\right)^{2(K_c+K_{sf}^{-1})-2}\right], \nonumber
\end{align}
with
\begin{align}
  c^T_\mu&= \frac{2K_c}{\chi_0 \pi^4 a_0^2}
           \cos\left(\frac{\pi(K_c+K_\mu)}{2}\right)
           B^2\left(\frac{K_c+K_\mu}{2},1-K_c-K_\mu\right), \nn 
{\overline{c}^T_{sf}} &= \frac{2K_c}{\chi_0 \pi^4 a_0^2}
                     \cos\left(\frac{\pi(K_c+K_{sf}^{-1})}{2}\right)
                     B^2\left(\frac{K_c+K_{sf}^{-1}}{2},1-K_c-K_{sf}^{-1}\right). \nonumber
\end{align}
\end{widetext}

\subsubsection{RG-improvements}
As we described above, we also calculate the ``RG-improved'' expression to take into account higher order corrections. Formally, iteration of the RG procedure for the optical conductivity leads to the scaling relation
\be
\sigma(\omega,T;\{ g\})= \exp(l)\,\sigma\Big(\,\exp(l) \omega,\, \exp(l) T;\,\{g(l)\}\Big), \nonumber
\ee
which in terms of the memory function is
\be
\sigma(\omega,T)=\frac{i\chi_0}{\omega}\frac{1}{1+m[\exp(l)\omega,\exp(l)T;\{g(l)\}]}, \nonumber
\ee
where we have defined $m(\omega,T;\{g\})=M(\omega,T;\{g\})/\omega$. In this expression, we replace the bare couplings and exponents $\{g\}=\{g_\mu, K_\mu\}$ by their scale-dependent analogues and identify the RG-scale and frequency via $l=\ln(\max(\omega,T)/\Lambda)$ with $\Lambda=v/a_0$ being the UV cutoff. The lowest energies we can reach are then determined by the scale $l^*$ at which the couplings become of order one, where we consequently stop the RG flow. This leads to $\omega_{low}, T_{low}\sim\Lambda\exp(-l^*)$. In the case of the exactly solvable sine-Gordon model, it was shown that this RG-improved perturbation theory approximates the exact optical conductivity very well~\cite{controzzi2001optical}.

\subsection{Results}

\begin{figure}[t]
\includegraphics[width=0.48\textwidth]{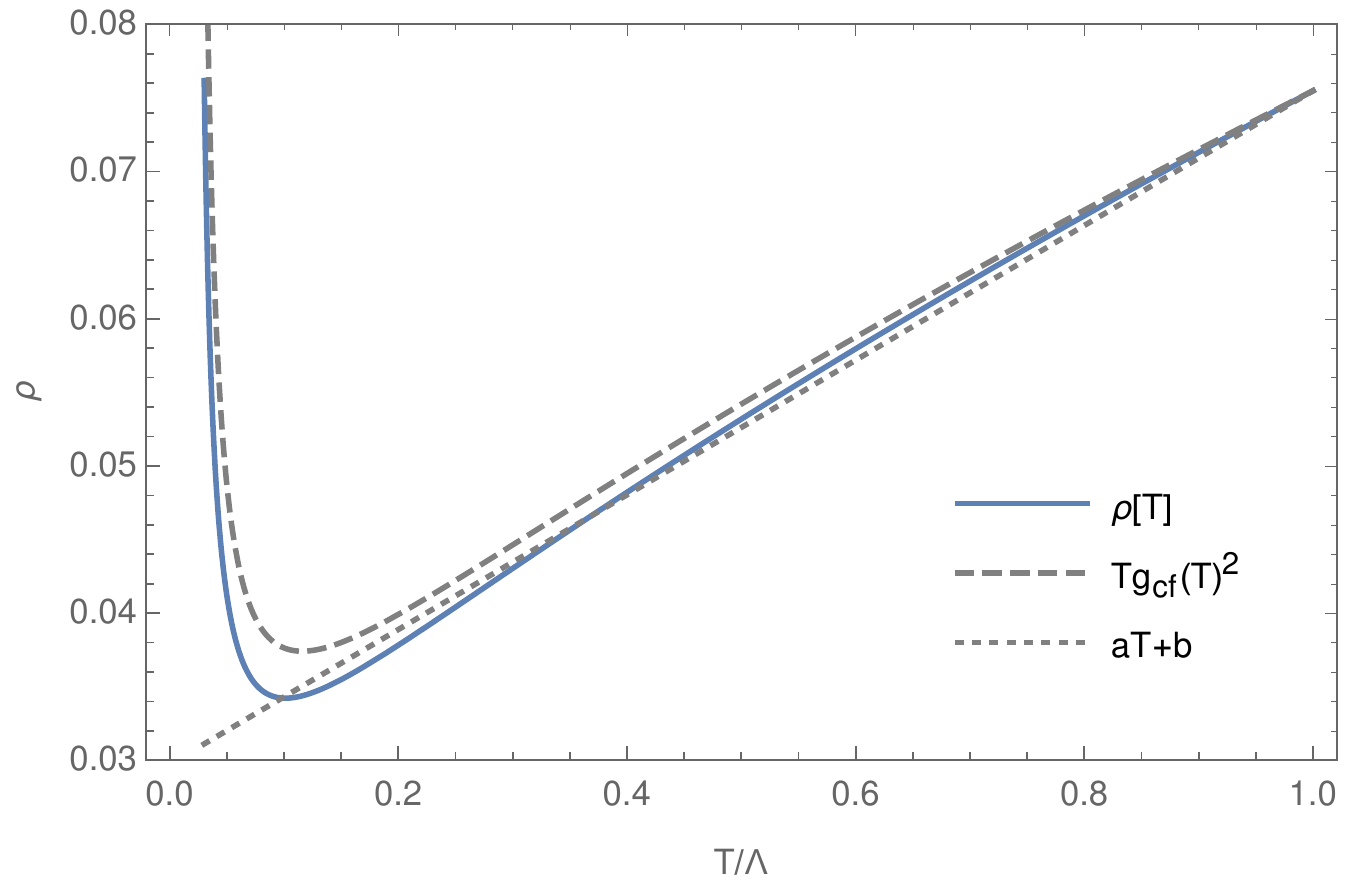}
\caption{The RG-improved resistivity as a function of temperature, $\rho(T)=\text{Im}\,M(\omega=0,T)/\chi_0$. The gray dashed line represents the approximation $\propto Tg^2(T)$ with $g\in\{g_{cf},g_{cs},g_{csf}, g_{\overline{csf}}\}$ and $g_{cf}$ chosen as an illustrative example (similarly good agreement is seen for any choice of $g$). The dotted line is a linear fit, $aT+b$ with $a,b$ fit parameters. The temperature (frequency) scale is set by the UV cutoff of our theory, which corresponds to the energy where the dispersion can no longer be approximated as linear.}
\label{fig:resT}
\end{figure}

Let us first discuss the two different limits, $\omega\ll T$ and $T\ll\omega$. From the extrapolation of $\sigma(\omega)$ to  $\omega\rightarrow0$, we can determine the dc resistivity
\be
\rho(T)=\frac{1}{\re\,\sigma(\omega=0,T)} = \frac{1}{\chi_0}\text{Im}\,M(\omega=0,T).
\ee
The RG-improved memory function scales like $M(T)\propto T g^2$ in this limit. As a result, we can approximate our calculated dc resistivity very well by 
\be
\rho(T)\propto g^2(T)\, T 
\ee
for any of the couplings $g\in\{g_{cf},g_{cs},g_{csf}, g_{\overline{csf}}\}$; the illustrative example of $g_{cf}$ is shown in Fig.~\ref{fig:resT}. Starting from high temperatures the resistivity first decreases upon lowering the temperature, before rapidly increasing when the temperature approaches the gap scale (where all the coupling start to grow). We expect this increase to cross over to an exponential increase below the gap, as there the number of carriers becomes exponentially small.  At intermediate-to-high temperatures, we also show that a resistivity growing linearly in the temperature fits the full expression well. Indeed, a similar behavior was predicted in Ref.~\onlinecite{rice2017umklapp} using qualitative arguments. As explained there, if the linear-in-temperature component is complemented with the contribution from the nodal Fermi liquid (not contained in our mapping), one obtains behaviour consistent with the resistivity observed in many cuprates, $\rho(T)\propto T[\exp(-\alpha/T)+\beta/T]^{-1}$. This gives Fermi liquid-like scaling with temperature at small $T$, which becomes linear at higher temperatures.

\begin{figure}[t]
\includegraphics[width=0.48\textwidth]{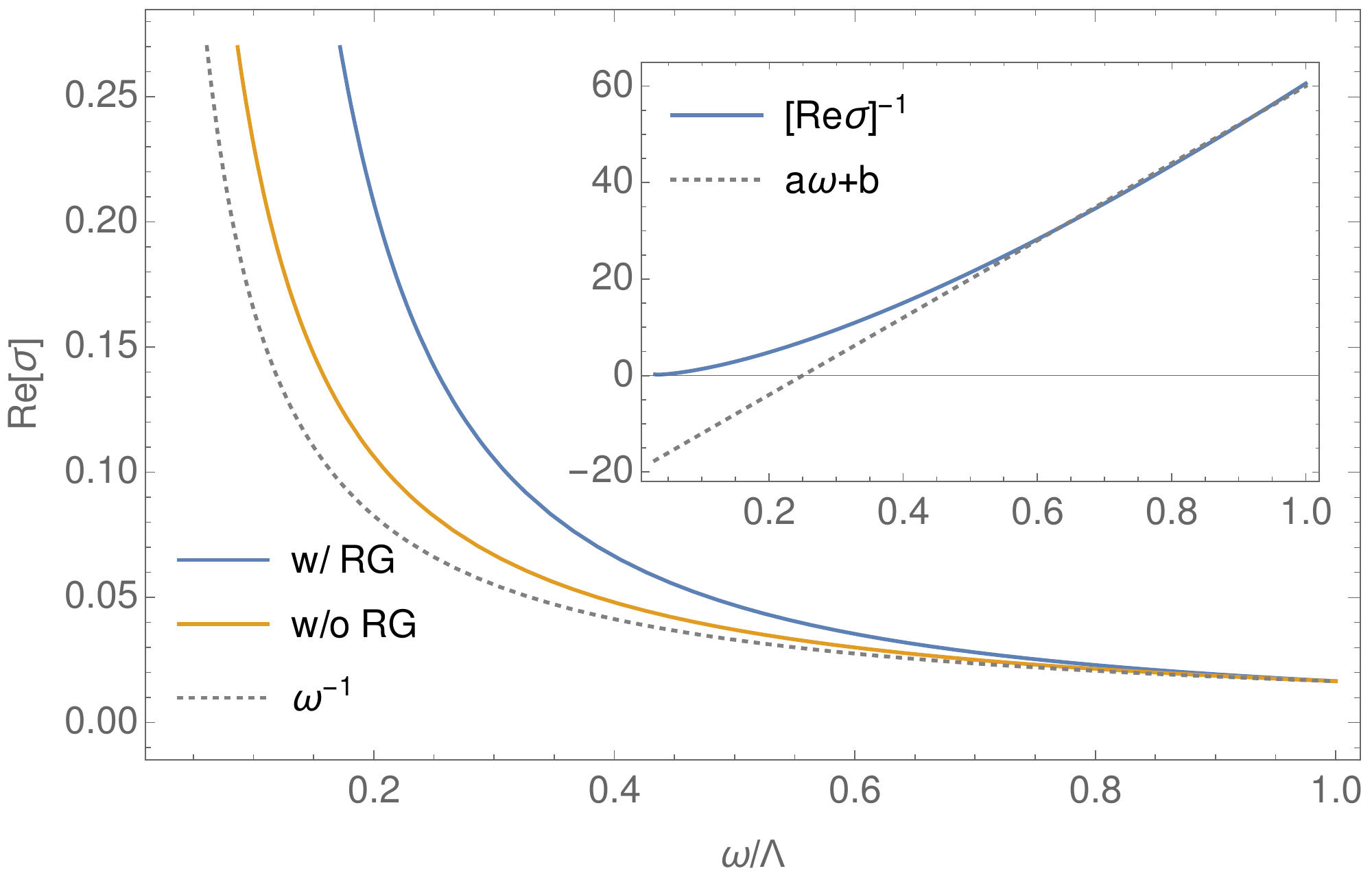}
\includegraphics[width=0.48\textwidth]{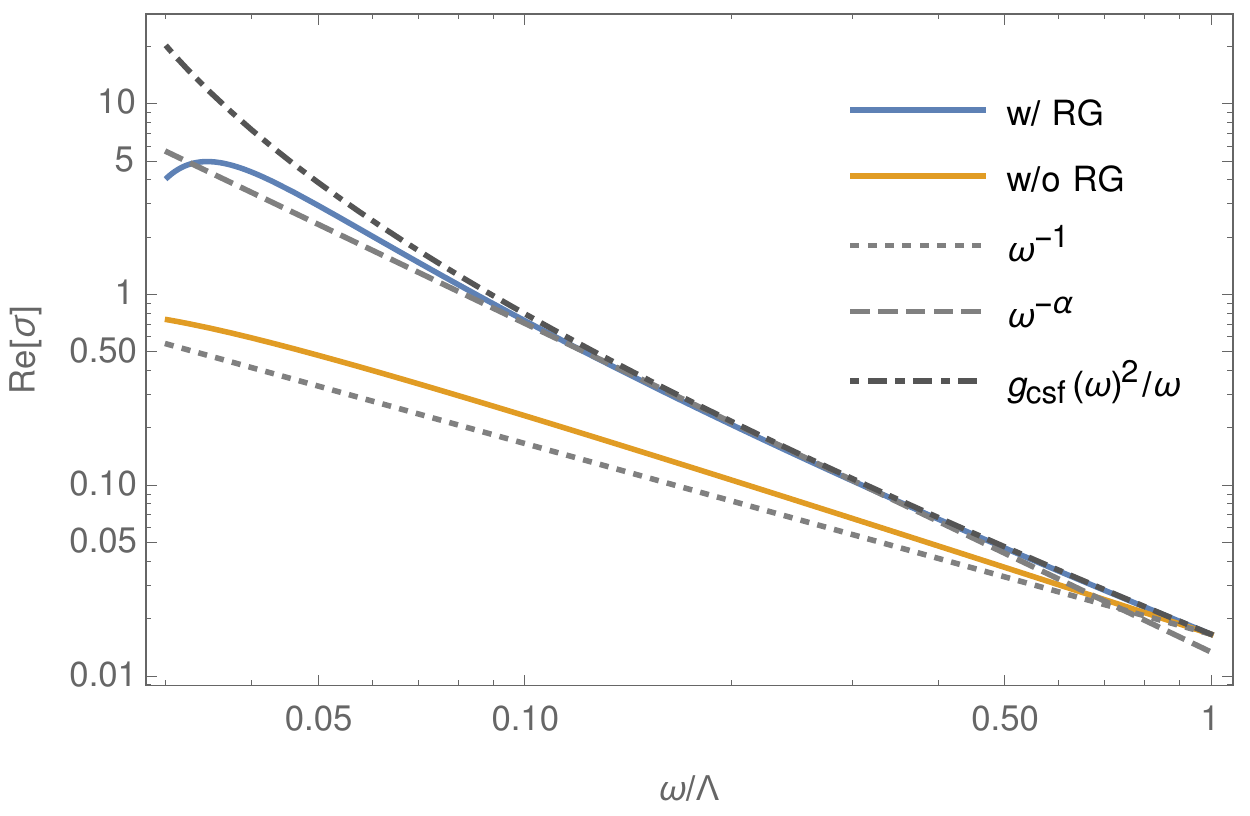}
\caption{The RG-improved optical conductivity $\sigma(\omega)$ in the zero-temperature limit $\omega\gg T$. We also show the bare expression (w/o RG) and a $1/\omega$-tail (dotted line) for comparison. In the inset, the inverse of the optical conductivity is approximated by a linear function of frequency. The lower panel presents the same results on a log-log scale. The RG-improved result approximately falls off like a power law $\omega^{-\alpha}$ with a non-universal exponent, e.g., $\alpha\approx1.70-1.97$ for $\gamma=0.15v-0.3v$. The contribution from the spin-flavor sector $\re\,\sigma(\omega)\propto g^2_{csf}(\omega)/\omega$ (gray, dash-dotted) is shown, with similar behavior seen for any choice of $g\in\{g_{cf},g_{cs},g_{csf}, g_{\overline{csf}}\}$.}
\label{fig:cond0}
\end{figure}

The optical conductivity as a function of frequency in the limit $T\rightarrow0$ is presented in Fig.~\ref{fig:cond0}. In this limit, the RG-improved memory function becomes $M(\omega)\propto \omega g^2$, such that the optical conductivity becomes
\be
\re\,\sigma(\omega,T=0)\approx\chi_0 \frac{\text{Im}\,M}{\omega^2} \propto \frac{g^2(\omega)}{\omega}.
\ee
Comparing the contributions from different couplings $g\in\{g_{cf},g_{cs},g_{csf}, g_{\overline{csf}}\}$, we find again that they all lead to a similarly good approximation. This reflects the fact that all couplings are of the same order and so any difference is essentially invisible on logarithmic scale.

The effect of the RG flow of the couplings is clearly visible when comparing the RG-improved result to the bare calculation, Fig.~\ref{fig:cond0}. Here, we also include the running of the Luttinger parameters, which is frequently neglected because of their weak scale dependence (Fig.~\ref{fig:flow}). We find that such approximation is justified at high frequencies, but changes the behavior of the optical conductivity at intermediate-to-low frequencies.
 
At high-to-intermediate frequencies we find that the optical conductivity is best approximated by a frequency dependence of the form
\be
\re\,\sigma(\omega)\sim \omega^{-\alpha}, \quad \alpha>1
\ee
as becomes apparent on a log-log scale (see the lower panel of Fig.~\ref{fig:cond0}). The exponent $\alpha$ is non-universal and depends on the initial (bare) interactions. Numerically, we obtain $\alpha\approx1.70-1.97$ for bare $\gamma=0.15v-0.3v$.

At smaller frequencies, the gap formation influences optical conductivity, which starts to deviate from the power-law scaling. This is driven by the RG flow to strong coupling. We can only observe the crossover regime within our formalism, as we are limited to frequencies sufficiently above the excitation gap. At low frequencies, we expect a suppression of the optical conductivity by the excitation gap~\cite{konik2001exact}. To capture the correct form of the singularity around the optical gap, a full knowledge of interacting matrix elements of the current operator is needed in addition to the gap formation~\cite{konik2001exact}. Furthermore, we have accounted only for contributions from the antinodal region; after the formation of a gap in the antinodal, we would need to include the contribution to the optical conductivity from the nodal quasiparticles. Here we expect a Drude-like behavior above the superconducting transition due to the Fermi-liquid character of the nodal quasiparticles.

\begin{figure}[t]
\includegraphics[width=0.48\textwidth]{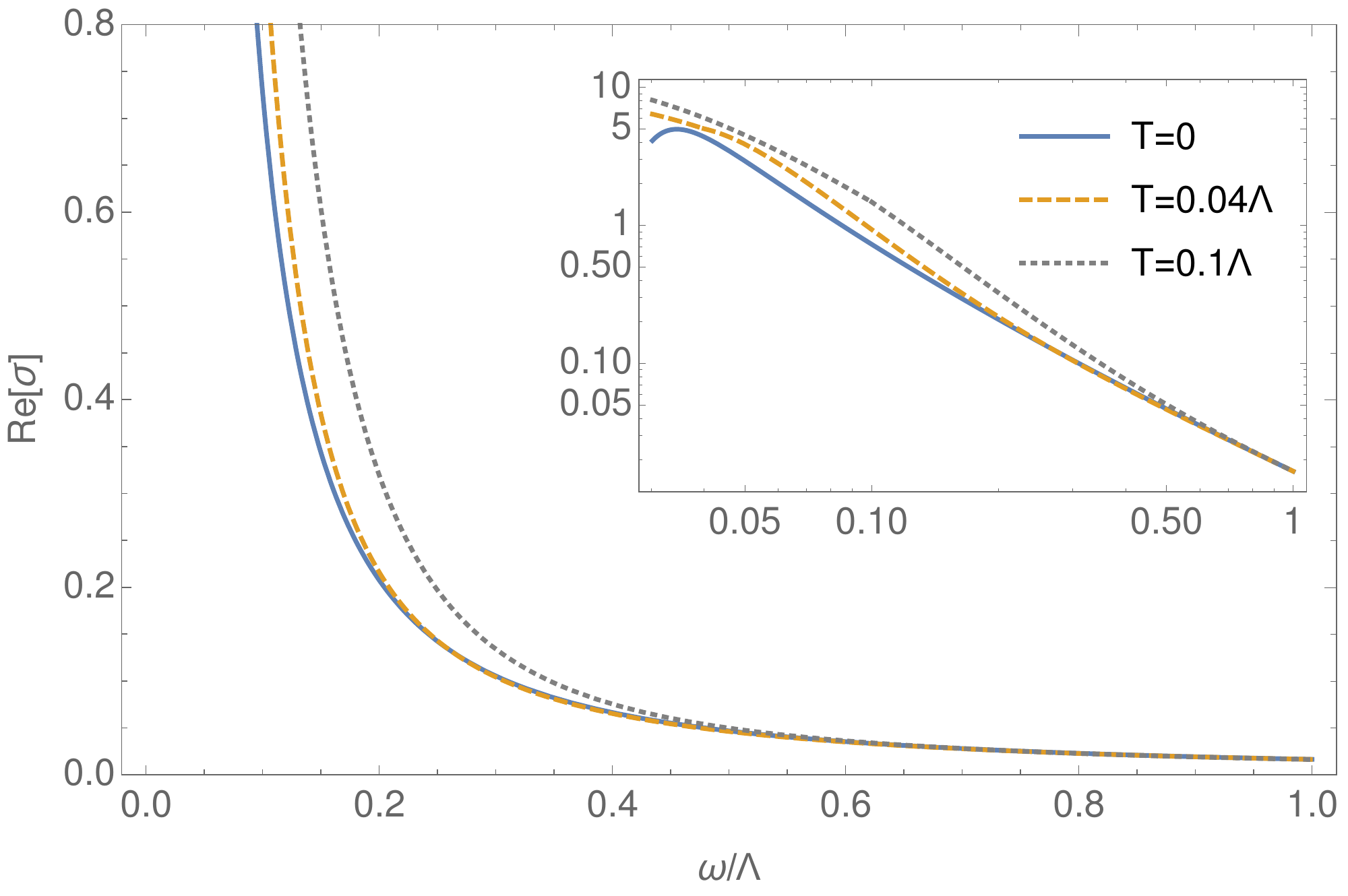}
\caption{The RG-improved optical conductivity $\sigma(\omega)$ for different temperatures $T=0,\,0.04\Lambda,\,0.1\Lambda$. The inset shows the same data on a log-log scale. A kink appears at $\omega=T$ due to our identification of the RG scale with $\ln(\max(\omega,T)/\Lambda)$.}
\label{fig:condT}
\end{figure}

For finite temperature and frequency (Fig.~\ref{fig:condT}), the high-frequency behavior follows that of the optical conductivity at $T=0$, with the proviso that at intermediate frequencies it increases more steeply. At low frequencies, this increase is slowed once again due to the RG flow of the couplings and Luttinger parameters, and it is cut when $\omega$ becomes smaller than $T$. The lower the temperature, the more the curve approaches the zero-temperature limit and becomes almost indistinguishable from it at the lowest temperatures we can reach $T_{low}\sim\Lambda \exp(-l^*)$. 

\begin{figure}[t]
\includegraphics[width=0.23\textwidth]{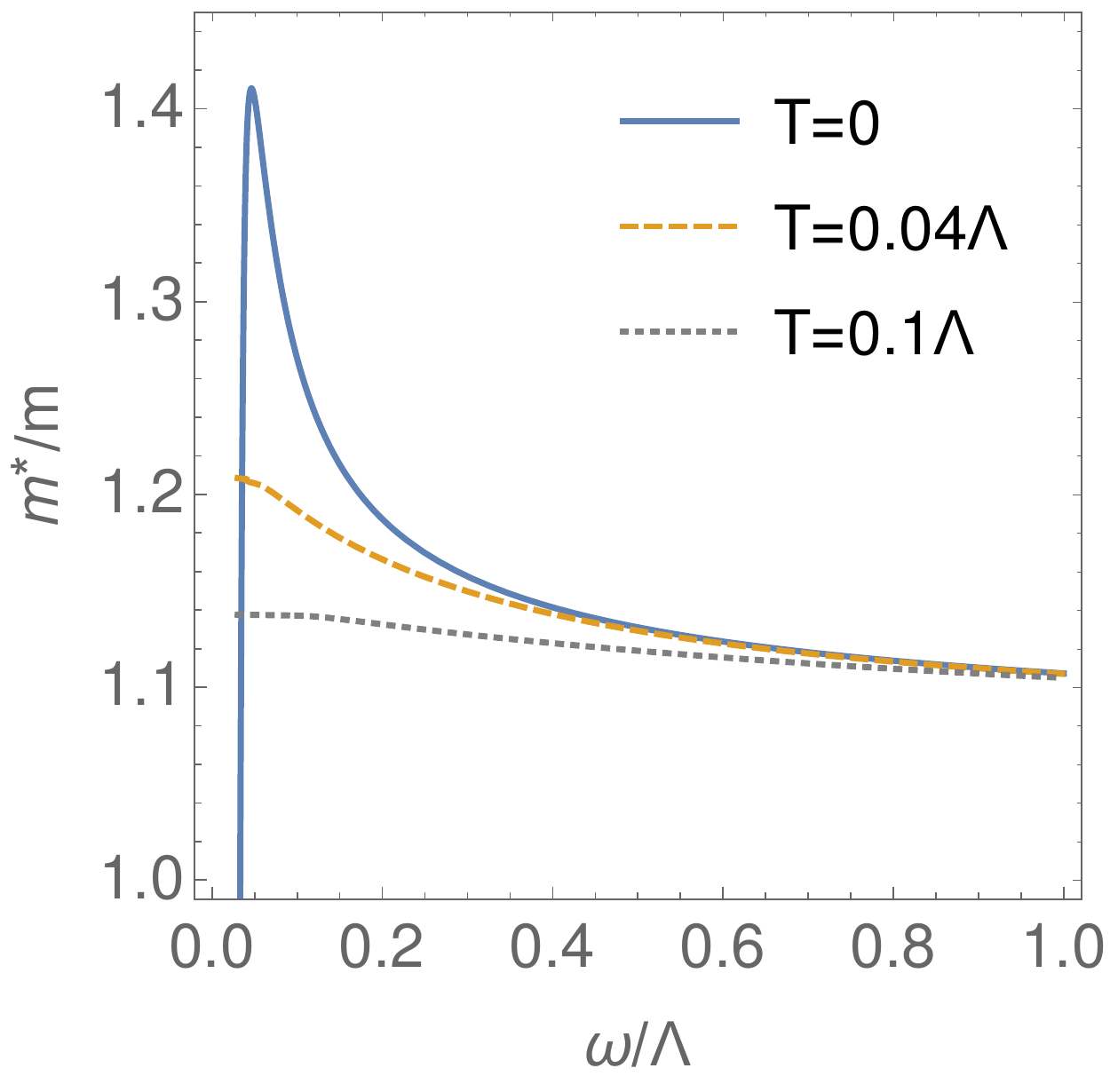}
\includegraphics[width=0.23\textwidth]{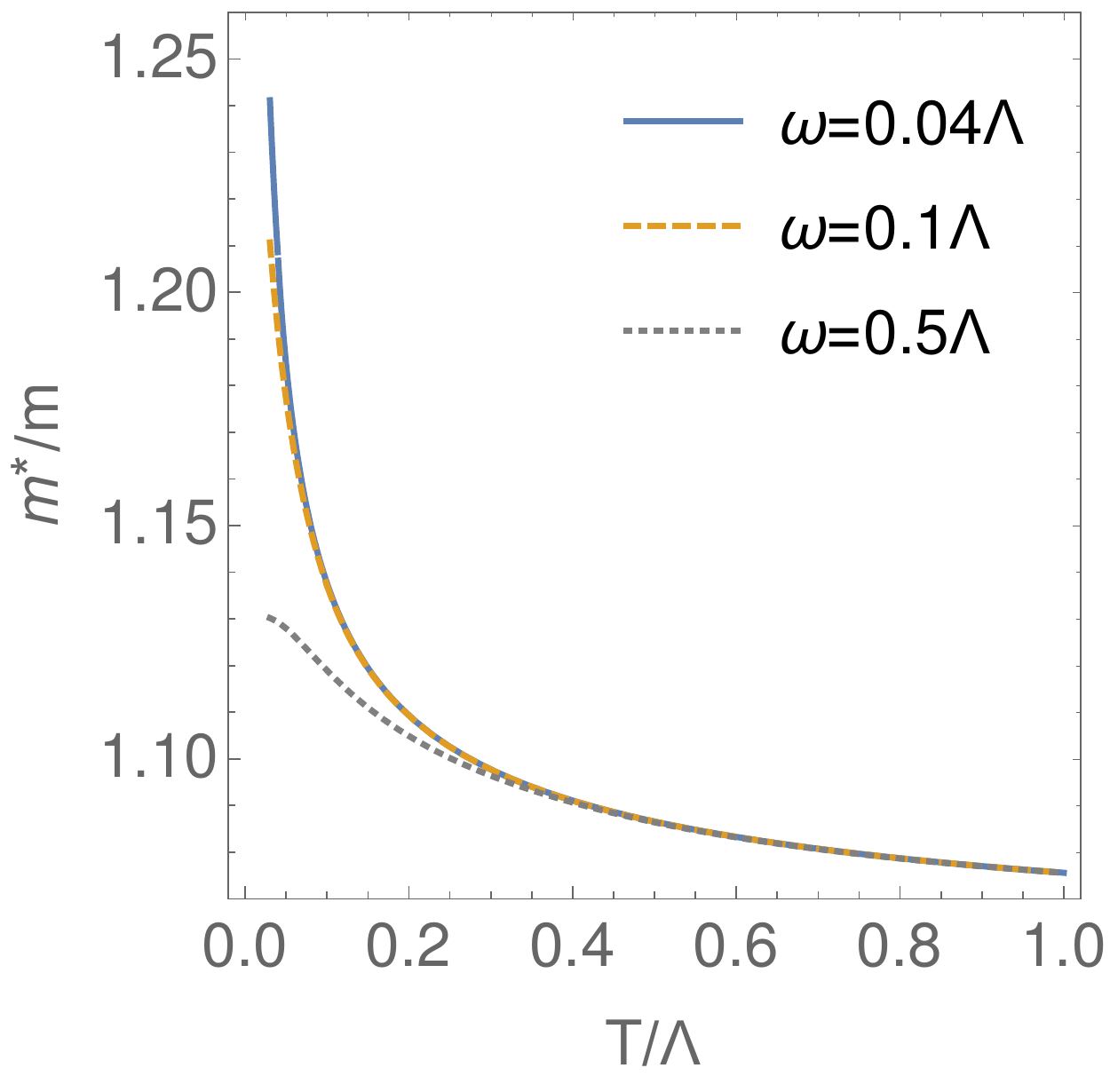}\\
\includegraphics[width=0.23\textwidth]{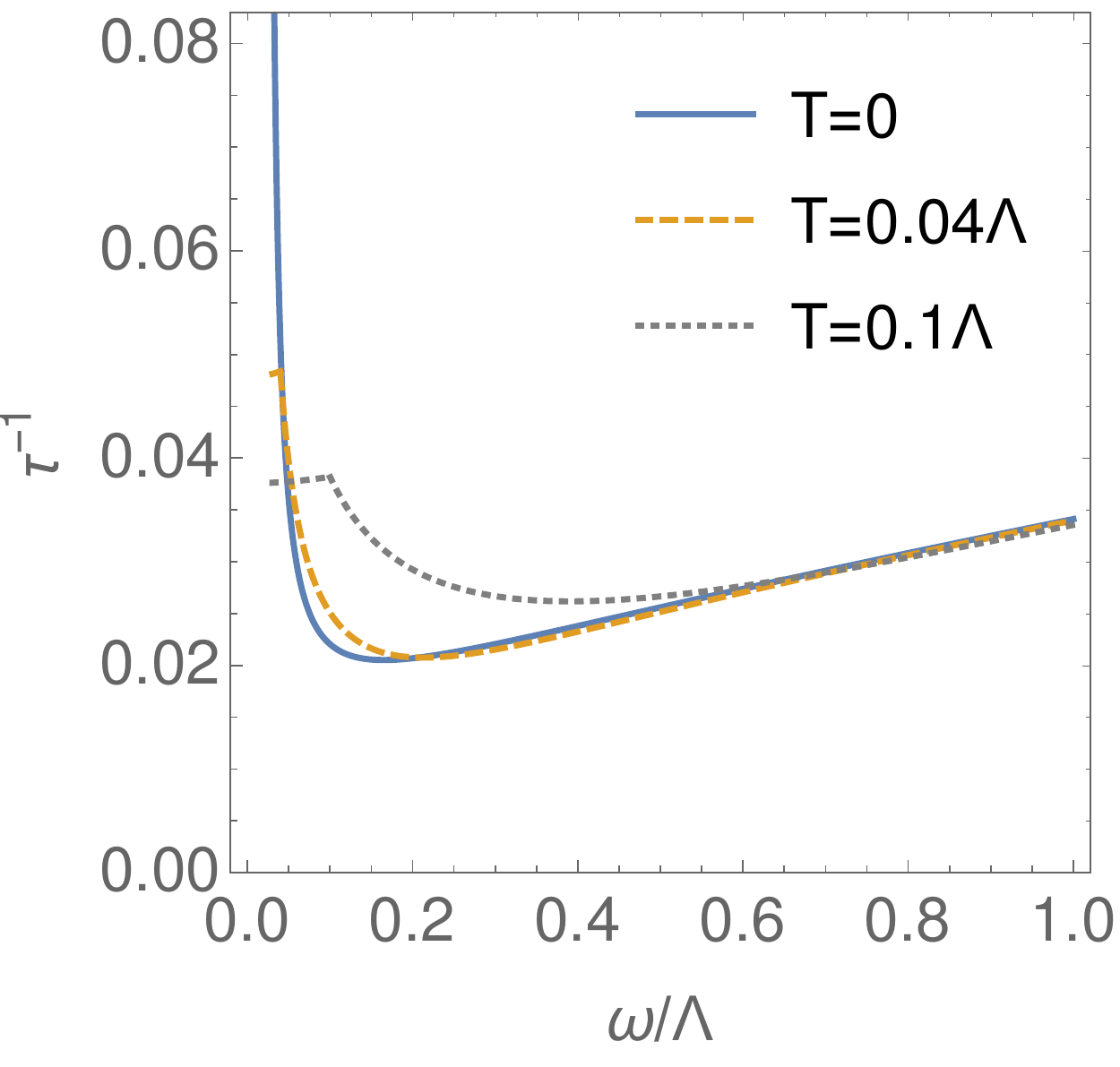}
\includegraphics[width=0.23\textwidth]{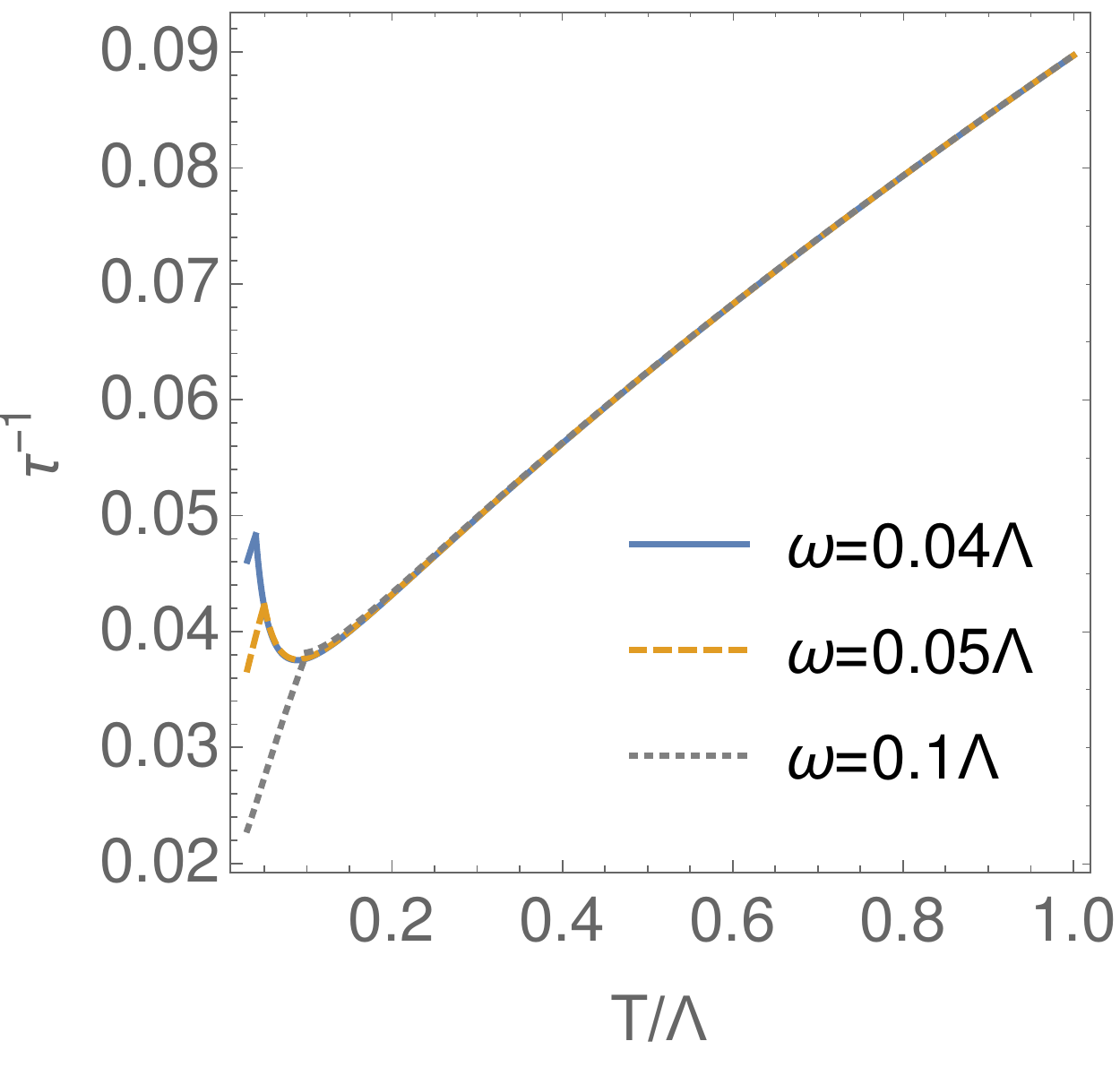}
\caption{Mass renormalization factor and relaxation rate as function of frequency and temperature. The abrupt kinks in the scattering rate come from the artificially nonanalytic switch in $l=\ln(\max(\omega,T)/\Lambda)$ }
\label{fig:mass}
\end{figure}

To make contact with the commonly used generalized Drude model (see, e.g., Ref.~\onlinecite{gotze1972homogeneous,puchkov1996pseudogap}), we show the generalized dynamical relaxation rate $1/\tau$ and mass renormalization factor $m^*/m$ in Fig.~\ref{fig:mass}. They are obtained from the memory function by
\begin{align}
\frac{1}{\tau(\omega,T)}=\text{Im}\,M(\omega,T), \quad \frac{m^*}{m}=1+\frac{\re\,M(\omega,T)}{\omega}
\end{align}
The mass renormalization factor remains close to one at higher frequencies or temperatures, but increases close to the gap opening where correlations become stronger. At zero temperature, we observe a maximum in the frequency dependence of the mass renormalization before the limiting value $\omega_{low}$ is reached. At higher temperature, the maximal value of the mass renormalization decreases.

The temperature dependence of the scattering rate for different fixed frequencies follows the approximately linear behavior of the zero-frequency limit shown earlier for the dc resistivity. However, the divergence at low temperatures is cut when the frequency exceeds the temperature $\omega>T$. Similarly, at fixed temperature the frequency-dependent scattering rate is approximately a linear at high frequency, and grows strongly at low frequencies. The low-frequency increase is suppressed at higher temperatures, and is also cut when $T>\omega$. For our calculations the cuts appears nonanalytic, but this is an artifact from our use of $\max(\omega,T)$ in the identification of the RG-scale with frequency or temperature. When an analytic identification is used, We expect this cut to become smoother.

The behavior of the generalized mass renormalization and scattering rate is in  qualitative agreement with measurements of the memory function in underdoped cuprates~
\cite{puchkov1996pseudogap,mirzaei2013spectroscopic} at frequencies/temperatures above the pseudogap scale. However, let us note again, that at low frequencies or temperatures the contribution from nodal regions also has to be taken into account.

Finally, we compare our result to measurements of the optical conductivity in underdoped YBa$_2$CuO$_{6.6}$ for light polarized along the $a-$axis~\cite{homes1999effect}. According to our theory there is a $(T,\omega)$ threshold below which the conductivity is a sum of independent nodal and antinodal contributions. This threshold is the energy at which flat portions of the Fermi surface form in the antinodal regions.
To isolate the nodal contribution, we adopt the Fermi liquid expression for it:
\begin{align}
\re\,\sigma(\omega)&=A\frac{1/\tau}{\omega^2+1/\tau^2}\\
\frac{1}{\tau} &= \Gamma_0+ B\Big[\omega^2 + (2\pi T)^2\Big],
\end{align}
and we extract parameters $A,B,\Gamma_0$ from experimental data under the assumption that at $T=70K$ the antinodal region is completely gapped and does not significantly contribute to the optical conductivity.

\begin{figure}[t]
\includegraphics[width=0.48\textwidth]{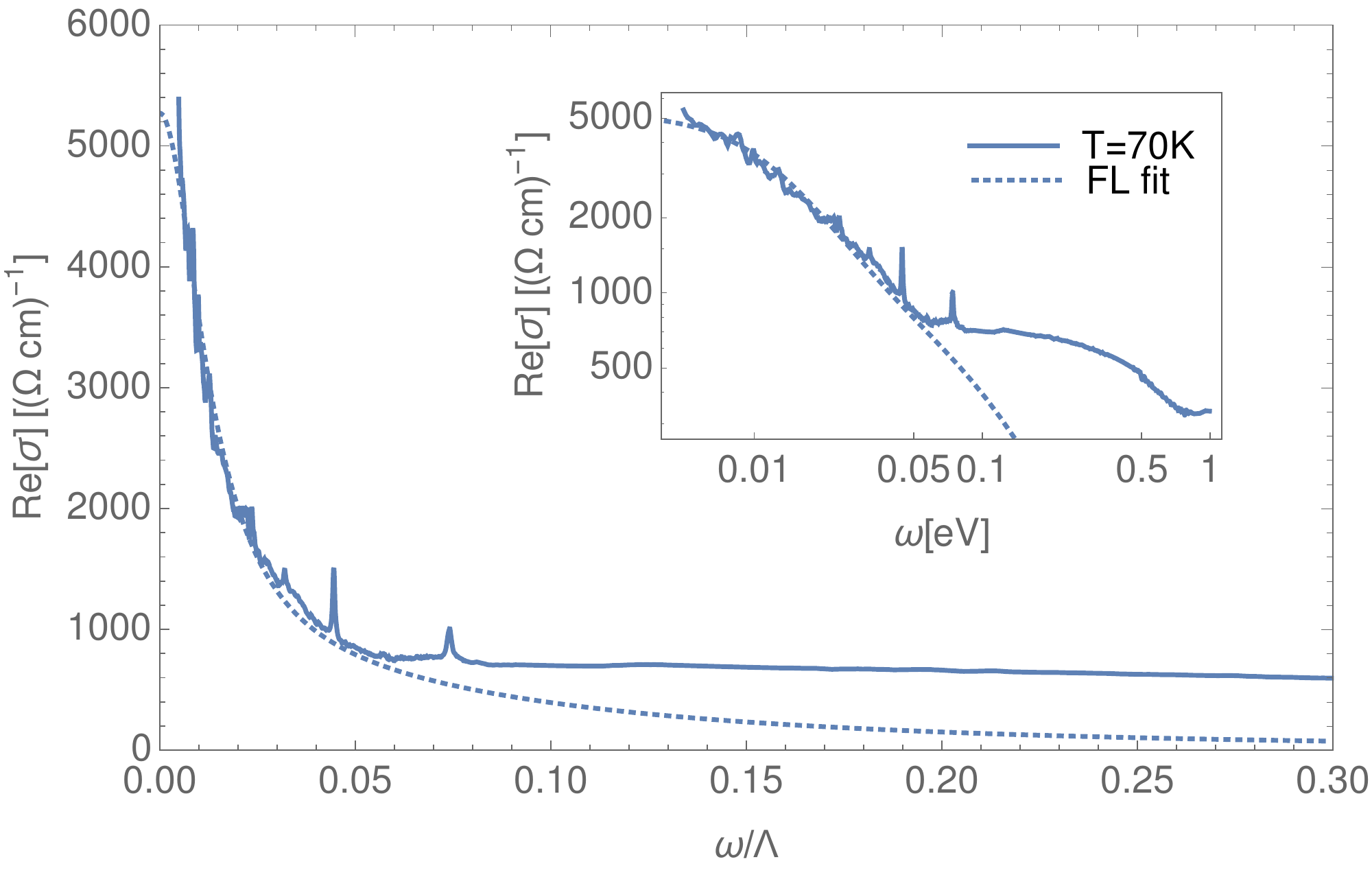}
\caption{The Fermi liquid fit to the experimental data~\cite{homes1999effect} for optical conductivity in underdoped YBCO  ($T_c \simeq 57$\,K) at $T=70$\,K. }\label{fig:exp1}
\end{figure}

\begin{figure}[t]
\includegraphics[width=0.48\textwidth]{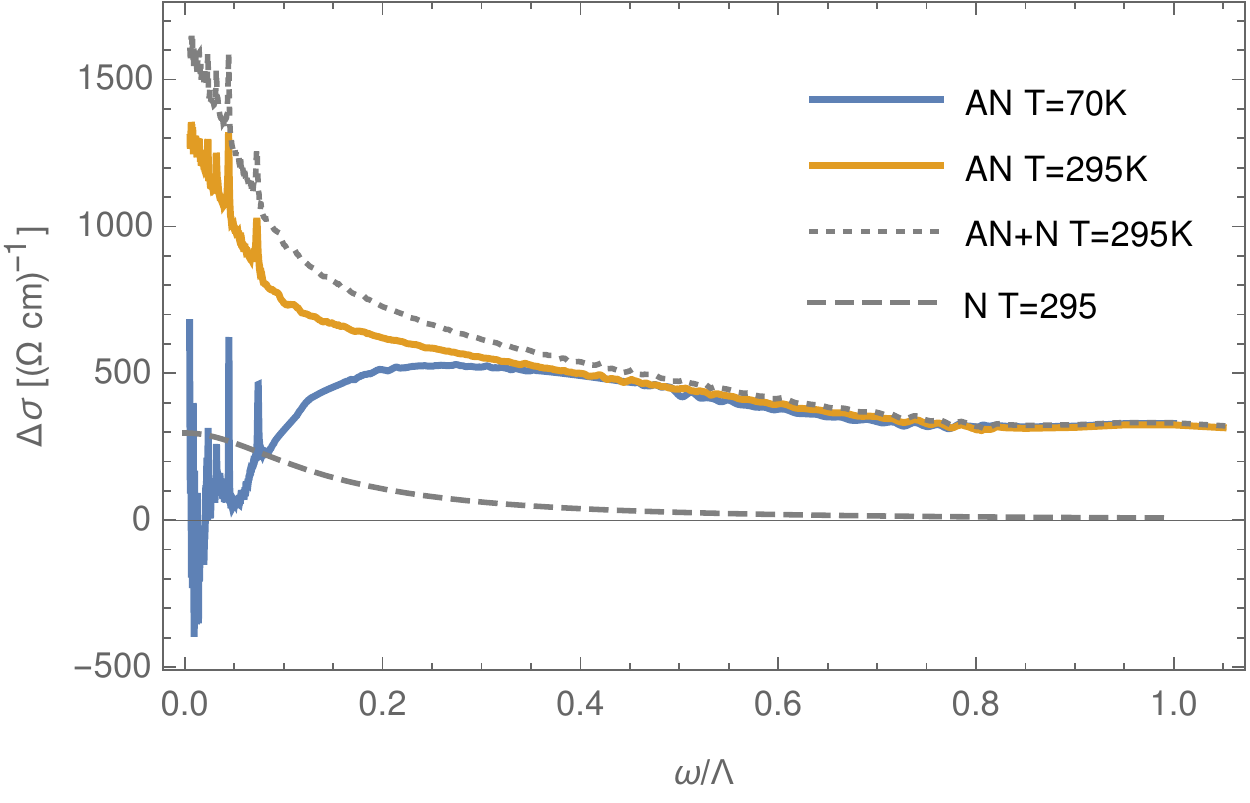}
\caption{The nodal (N) and antinodal (AN) contribution to the optical conductivity in underdoped YBCO at $T =70$\, K and $T=295$\,K obtained by subtraction of the Fermi liquid contribution.}\label{fig:exp2}
\end{figure}

Having determined the fit parameters, we subtract the Fermi liquid contribution to the optical conductivity at $T=295$\,K,  revealing the antinodal contribution (under our assumptions). We obtained $\Gamma_0=0, A=80.508$ eV/($\Omega$cm) and $B=10.675$ eV$^{-1}$.  Figure~\ref{fig:exp2} indicates the consistency of our original assumption~\cite{rice2017umklapp}: above the pseudogap the conductivity is dominated by the antinodal regions.
Although our theory provides an explanation of why spectral weight is found at higher frequencies, we have difficulty fitting the power law of the high-frequency tail we find. However, as we have noted, this power law is non-universal and quantitative statements about it are beyond our approach. There are several potential reasons for the mismatch: coupling between the ladders $(R,L)$ and $(\bar R, \bar L)$ in Fig.~\ref{Fig:SpinFermionBZ} or other excitations beyond the 1D picture may contribute at these frequencies. Furthermore at high frequencies and temperatures above the pseudogap, our assumption of a flat antinodal Fermi surface could become invalid.

\section{Conclusions}

We have obtained analytical results for the conductivity of the spin-fermion model with flat portions of the Fermi surface at temperatures and frequencies above the pseudogap.  Such a problem can be mapped to a model of 1D half filled ladder.
The 1D-like behavior emerges in the limit where the Fermi surface around the hotspots becomes increasingly nested, which was conjectured to be energetically favorable for strong coupling because the nesting leads to the formation of a gap\cite{tsvelik2017ladder}. We calculated the corresponding RG evolution of the ladder model, which flows to strong coupling, with all excitations developing a gap (the pseudogap). We showed that the system possesses SO(5)$\times$SO(3) symmetry in the relevant intermediate energy regime, where excitations can be classified as in the well-known SO(8) symmetric limit. The low-energy effective theory describes the d-Mott phase, where there is a spin gap and short-ranged d-wave pairing correlations.

Our calculation of the optical conductivity falls in a different regime of the spin-fermion model than previously studied, because it assumes full nesting of the hotspots, together with non-zero coupling to the collective spin excitations. At the same time, it differs from previous theoretical studies of the optical conductivity in ladder system because the ladder is  away from the integrable  SO(8) limit where exact results can be obtained. We argued that in the considered frequency and temperature regime, the  contribution of the antinodal regions to the conductivity $\sigma_{an}$ is determined by umklapp processes (Fig.~\ref{fig:Umklapps}). By combining a perturbative memory function approach with one-loop renormalization group, we show that the optical conductivity scales like $\re\,\sigma_{an}\propto g^2(\omega)/\omega$ for zero temperature. The dc resistivity follows from the zero frequency limit, giving $\rho_{an}(T) \propto g^2(T) T$. The coupling $g$ determines the strength of the umklapp processes and its frequency (temperature) dependence is given by its scale dependence as obtained from the RG flow. Our results for the dc resistivity support the qualitative picture presented in Ref.~\onlinecite{rice2017umklapp}: 
 above the pseudogap the conductivity is increasingly dominated by the \textit{antinodal regions} and the temperature dependence of the resistivity becomes approximately linear, which  reflects the effective one dimensionality of the system. 
 
In conclusion, we showed that the optical conductivity of the spin-fermion model (with parameters above the pseudogap phase) fits within the `two weakly-coupled fluids' picture of transport in the underdoped cuprates. In particular, the two fluids have different effective dimensionalities, with one describing the 2D Fermi-liquid of nodal electrons and the other a quasi-1D strongly correlated liquid around the hot spots. In future studies, it would be interesting to address the coupling between the two liquids to study the mutual effect that both components have on each other, e.g., with respect to Fermi-surface deformation and long-ranged superconductivity.

\label{sec:conc}

\acknowledgments
We acknowledge valuable discussions with Chris Homes who also acquainted us with his data on the infrared optical conductivity. We are also grateful to Andrey Chubukov, Gabriel Kotliar, Maurice Rice and John Tranquada for thoughtful comments and interest in the work. This project has received funding from the Alexander-von-Humboldt foundation (L.C.), the European Union's Horizon 2020 research and innovation program under grant agreement No.~745944 (N.J.R) and the U.S. Department of Energy, Office of Basic Energy Sciences, Contract No.~DE-SC0012704 (A.M.T., L.C.).

\bibliography{minbib}

\end{document}